\documentclass{article}

\usepackage[preprint,nonatbib]{neurips_2020}

\usepackage[utf8]{inputenc} 
\usepackage[T1]{fontenc}    
\usepackage{hyperref}       
\usepackage{url}            
\usepackage{booktabs}       
\usepackage{amsfonts}       
\usepackage{nicefrac}       
\usepackage{microtype}      

\usepackage[pdftex]{graphicx}

\usepackage[american]{babel}	
\usepackage{amsmath}            
\usepackage{bm}             
\DeclareMathOperator*{\argmax}{arg\,max}

\usepackage[capitalize,nameinlink]{cleveref} 

\usepackage[square,sort,comma,numbers]{natbib}
\usepackage{color}
\definecolor{tangoblue1}{rgb}{0.125,0.29,0.529}
\definecolor{tangoblue2}{rgb}{0.2,0.4,0.65}
\definecolor{tangoblue3}{rgb}{0.13,0.3,0.53}
\hypersetup{
     colorlinks   = true,
     citecolor    = tangoblue3,
     linkcolor    = tangoblue3}

\usepackage[font={normalsize},labelfont={color=tangoblue3}]{caption}
\DeclareGraphicsExtensions{.pdf, .png, .jpg, .eps}     
\graphicspath{{.}}         
\newcommand{\figwidth}{1.0\textwidth}
\usepackage{placeins}   

\newcommand{\beginsupplement}{%
        \setcounter{table}{0}
        \renewcommand{\thetable}{S\arabic{table}}%
        \setcounter{figure}{0}
        \renewcommand{\thefigure}{S\arabic{figure}}%
        \setcounter{section}{0}
        \renewcommand{\thesection}{S\arabic{section}}%
     }
\usepackage{multibib}
\newcites{SM}{Supplementary References}

\title{The interplay between randomness and structure during learning in RNNs}

\author{%
  Friedrich Schuessler
  \\
  Technion\\
  \texttt{schuessler@campus.technion.ac.il} \\
  \And
  Francesca Mastrogiuseppe\\
  Gatsby Unit, UCL\\
  \texttt{f.mastrogiuseppe@ucl.ac.uk} \\
  \And
  Alexis Dubreuil\\
  ENS Paris\\
  \texttt{alexis.dubreuil@gmail.com} \\
  \And
  Srdjan Ostojic\\
  ENS Paris\\
  \texttt{srdjan.ostojic@ens.fr} \\
  \And
  Omri Barak\\
  Technion\\
  \texttt{omri.barak@gmail.com} \\
}

\begin{document}

\maketitle

\begin{abstract}
Recurrent neural networks (RNNs) trained on low-dimensional tasks have been widely used to model functional biological networks. However, the solutions found by learning and the effect of initial connectivity are not well understood. Here, we examine RNNs trained using gradient descent on different tasks inspired by the neuroscience literature. We find that the changes in recurrent connectivity can be described by low-rank matrices, despite the unconstrained nature of the learning algorithm. To identify the origin of the low-rank structure, we turn to an analytically tractable setting: training a linear RNN on a simplified task. We show how the low-dimensional task structure leads to low-rank changes to connectivity. This low-rank structure allows us to explain and quantify the phenomenon of accelerated learning in the presence of random initial connectivity. Altogether, our study opens a new perspective to understanding trained RNNs in terms of both the learning process and the resulting network structure.
\end{abstract}

\section{Introduction}
Recurrent neural networks (RNNs) have been used both as tools for machine learning, and as models for neuroscience. In the latter context, RNNs are typically initialized with random connectivity and trained on abstractions of tasks used in experimental settings \cite{sussillo2014neural,barak2017recurrent,richards2019deep,yang2019task,sussillo2009generating,wang2018flexible,mante2013context,sussillo2015neural}. The obtained networks are then compared to both behavioral and neural experimental results, with the added advantage that the RNNs are more amenable to analysis than their biological counterparts \cite{sussillo2013opening}. Despite this advantage, the understanding of how RNNs implement neuroscience tasks is still limited. Open questions concern especially the relationship between the final connectivity and the task, and its formation through training.

Here, we examine the relation between the initial connectivity of the RNN, the task at hand, and the changes to connectivity through training. We use unconstrained gradient descent that can potentially alter the connectivity completely. However, evaluating nonlinear RNNs trained on several neuroscience-inspired tasks, we observe that the connectivity changes are small compared to the initial connectivity. We thus split the connectivity matrix $W$ at the end of training into the initial part $W_0$ and the changes $\Delta W$, writing
\begin{equation}
    W = W_0 + \Delta W \,.
\end{equation}
For all tasks we consider, we find that the training-induced connectivity structure $\Delta W$ is of low rank, despite the unconstrained nature of training used. This finding directly connects  gradient-based learning with a number of existing neuroscience frameworks based on low-rank aspects of connectivity \cite{hopfield1982neural,sussillo2009generating,eliasmith2004neural,mastrogiuseppe2018linking,rivkind2017local,logiaco2019model,tirozzi1991chaos,jaeger2004harnessing,barak2020mapping}. Despite the low-rank nature of the \textit{changes} to connectivity $\Delta W$, the initial, full-rank, random connectivity $W_0$ plays an important role in learning. Consistent with previous work \cite{sussillo2009generating,schoenholz2016deep}, we find that the initial connectivity accelerates learning. Moreover we show that the final, trained network relies on correlations between $\Delta W$ and $W_0$.

In the second part of our work, we analyze the mechanism behind these observations in a simplified and analytically tractable setting: nonlinear dynamics of learning in a linear RNN trained on a simple input-output mapping task. 
We show how the low-dimensional task structure leads to low-rank connectivity changes; importantly, the amplitude and geometry of these low-rank changes depend on the random initial connectivity. 
Our work reveals how this dependence accelerates learning and quantifies the degree of acceleration as a function of initial connectivity strength. 

Finally, we show that our results extend to real-world settings of an LSTM network trained on a natural language processing task, suggesting practical applications of our results.



\section{Training RNNs on low-dimensional tasks}
\paragraph{Tasks} 
We trained RNNs on three tasks inspired by the neuroscience literature. 
All tasks are characterized by a small number of input and output channels. 
The first task is a working memory task, in which the network receives pulses from two different input channels and needs to remember the sign of the last pulse in each channel independently \cite{sussillo2013opening}. 
The second task is a context-dependent decision task: The network receives two noisy signals, as well as one of two context inputs which indicates the relevant signal. After the input presentation, it needs to output whether the average of the relevant signal was positive or negative \cite{mante2013context}. 
The third task is a delayed-discrimination task \cite{romo1999neuronal} in which the network receives two positive pulses separated by a delay. After yet another delay, it needs to output which of the two pulses had the larger amplitude.
Based on their origin, we refer to the three tasks as "flip-flop" \cite{sussillo2013opening}, "Mante" \cite{mante2013context}, and "Romo" \cite{romo1999neuronal} task, respectively. 
For each task, we plotted a single trial for a successfully trained network in \cref{fig:all_tasks}\textbf{(a-c)}. Detailed parameters can be found in the supplementary. 

\begin{figure}
  \centering
  \includegraphics[width=\figwidth]{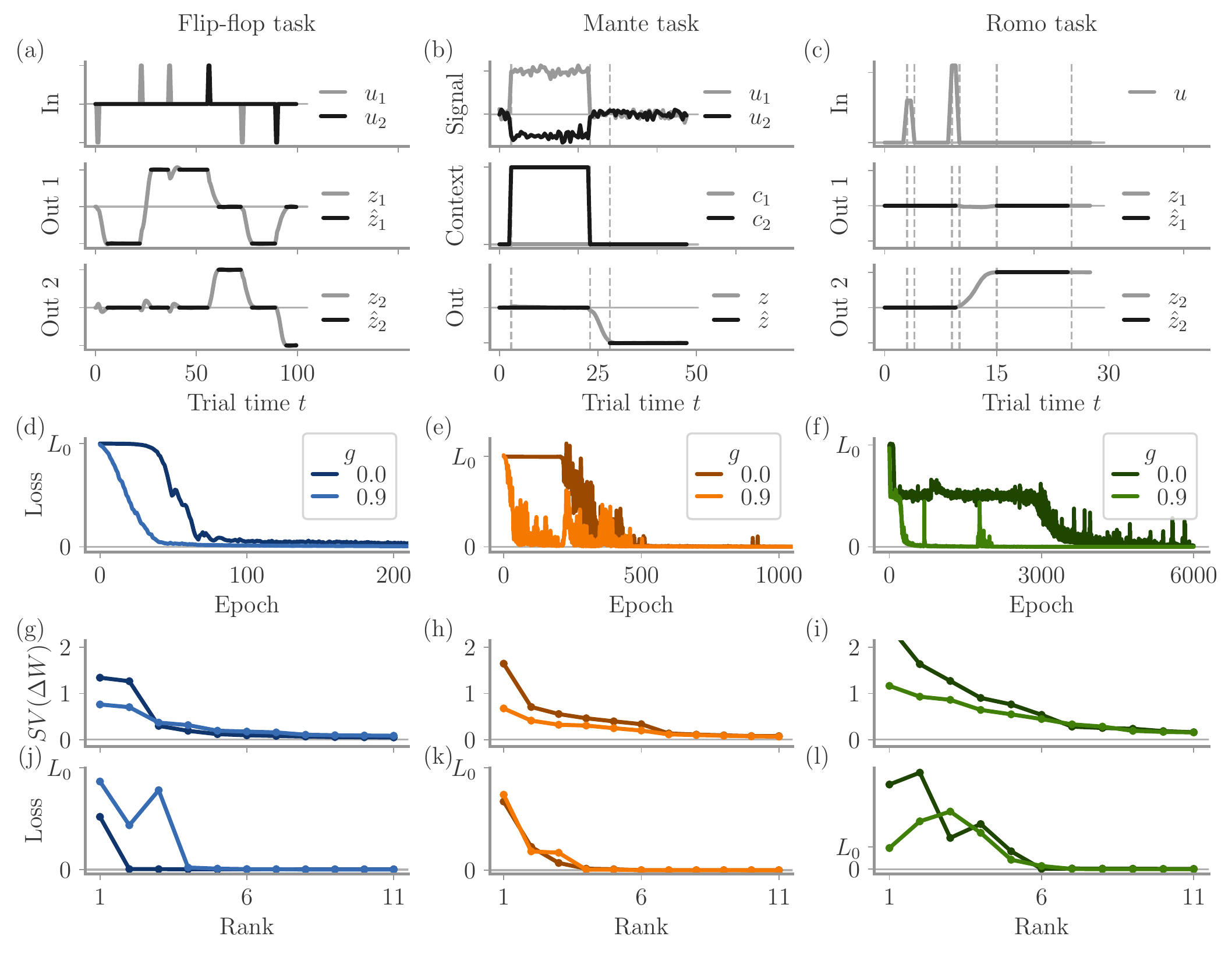}
  \caption{
  Learning dynamics in three different neuroscience tasks.
  \textbf{(a-c)}
  Task summary: inputs $u_i$, outputs $z_i$, and targets $\hat{z}_i$ for each task. Dashed lines indicate task phases. 
  \textbf{(d-f)} 
  Loss throughout training process for different initial connectivity strengths
  $g$. $L_0$ is the loss at the beginning of training for $g = 0$ ($L_0$ is different for different tasks). Note the different epoch numbers plotted. 
  \textbf{(g-i)}
  First 11 singular values of final connectivity changes $\Delta  W$.
  \textbf{(j-l)}
  Loss for truncated networks, where $\Delta W$ is replaced with the rank-$R$ approximation $\Delta W^{(R)}$. 
  Parameters: $N = 256$, learning rate $\eta = 0.05 / N$. 
  }
  \label{fig:all_tasks}
\end{figure}

\paragraph{RNN model} 
Each RNN model consists of $N$ neurons whose state vector evolves according to 
\begin{equation}
    \label{eq:dxdt_nonlin}
    \dot{\mathbf{x}}(t) = -\mathbf{x}(t) + W \phi(\mathbf{x}(t)) + \sqrt{N} \sum_{i=1}^{N_\mathrm{in}} \mathbf{m}_i u_i(t) \,.
\end{equation}
The recurrent input is given by the firing rate vector $\phi(\mathbf{x})$ multiplied by the weight matrix $W$. We use the element-wise nonlinearity $\phi = \mathrm{tanh}$. The network receives time-dependent inputs $u_i(t)$ through input vectors $\mathbf{m}_i$. The output is the projection of the firing rate onto readout vectors $\mathbf{w}_i$, namely 
\begin{equation}
    z_i(t) = \frac{\mathbf{w}_i^T\! \phi(\mathbf{x}(t)) }{\sqrt{N}}
    \quad \text{for $i$ in } \{1, \dots, N_\mathrm{out}\}  \,.
\end{equation}
We formulate target values $\hat{z}_i(t)$ during specific segments of the trial [see dark lines for output panels in \cref{fig:all_tasks}\textbf{(a-c)}]. 
The task determines the numbers $N_\mathrm{in}$ and $N_\mathrm{out}$ of input and output vectors.
For example, the Mante task requires four input vectors (for both signals and contexts) and a single output vector. 
We are interested in the behavior of large networks, $N >> 1$, while the dimension of the tasks is small, $N_\mathrm{in}, N_\mathrm{out} \sim \mathcal{O}(1)$. For the simulation, we chose $N$ to be large enough so that learning dynamics become invariant under changes in $N$ (see supplementary Fig. S1). 

\paragraph{Training and initialization} 
For training the RNNs, we formulated a quadratic cost in $z_i(t)$ and applied the gradient descent method ``Adam'' \cite{kingma2014adam} to the internal connectivity $W$ as well as to the input and output vectors $\mathbf{m}_i$, $\mathbf{w}_i$. 
Restricting the updates to $W$ or training with SGD impaired the convergence times but yielded similar results (not shown).  
The initial input and output vectors were drawn independently from $\mathcal{N}(0,\, 1/N)$. We initialized the internal weights as a random matrix $W_0$ with independent elements drawn from $\mathcal{N}(0,\, g^2/N)$. The parameter $g$ thus scales the strength of the initial connectivity. 

\paragraph{Learning dynamics in the absence of initial connectivity}
To understand what kind of connectivity arises during learning, we first looked at the simplest case without initial connectivity, $g = 0$. 
The loss curves indicate convergence for all three tasks [see darker lines in \cref{fig:all_tasks}\textbf{(d-f)}]. 
We analyzed the connectivity at the end of training by computing its singular values (SVs). For the flip-flop task, we found that the first two SVs were much larger than the remaining ones [\cref{fig:all_tasks}\textbf{(g)}]. To see whether the network utilizes this approximate rank-two structure, we replaced the changes $\Delta W$ with the singular value decomposition truncated at rank $R$, 
\begin{equation}
    \label{eq:dw_trunc}
    \Delta W^{(R)} = \sum_{r=1}^R s_r \mathbf{u}_r \mathbf{v}_r^T \,.
\end{equation}
Note that we keep the initial connectivity $W_0$. The loss after truncation indeed drops to zero at rank 2 [\cref{fig:all_tasks}\textbf{(j)}]. A similar situation is observed for the Mante and Romo tasks, see \cref{fig:all_tasks}\textbf{(h, k)} and \textbf{(i, l)}, respectively. 
Although for these tasks the SVs drop more slowly, the first six SVs are discernibly larger than the remaining tail; the truncation loss drops to zero at rank 4 and 6, respectively. 
In sum, we observe that for $g = 0$, training via gradient descent yields an effective low-rank solution for all three tasks.

\begin{figure}
  \centering
  \includegraphics[width=\figwidth]{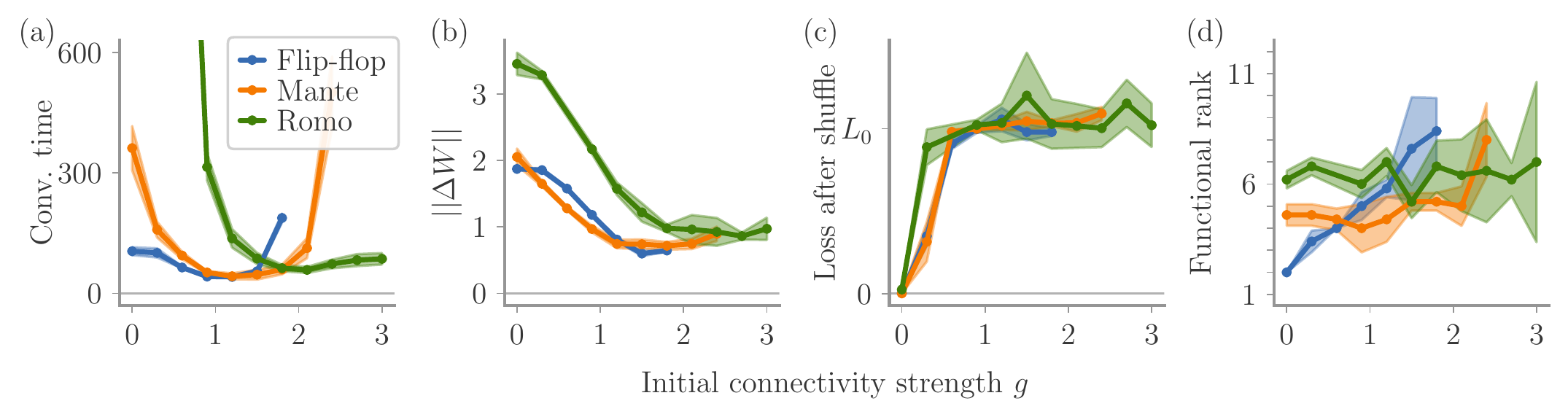}
  \caption{
  Dependence of learning dynamics on initial connectivity strength $g$ in the three tasks.
  Lines and shades indicate mean and standard deviation of five independent simulations for each $g$, respectively. 
  \textbf{(a)} 
  Number of epochs at which the loss falls below 5\% of $L_0$. 
  \textbf{(b)} 
  Frobenius norm of $\Delta W$ at the end of training.
  \textbf{(c)} 
  Loss for shuffled initial connectivity, so that the full network connectivity is given by $W_0^\mathrm{shuffle} + \Delta W$.
  \textbf{(d)}
  Rank $R$ at which the loss of the network with rank-truncated connectivity $\Delta W^{(R)}$ drops below 5\% of the initial loss $L_0$.
  }
  \label{fig:learning_time_min_rank}
\end{figure}

\paragraph{Effects of initial connectivity on learning dynamics and connectivity}
The loss-curves in \cref{fig:all_tasks}\textbf{(d-f)} indicate a strong influence of the initial connectivity strength $g$ on the training dynamics (lighter colors for $g = 0.9$). We observe that learning becomes faster and smoother with initial connectivity. In \cref{fig:learning_time_min_rank}\textbf{(a)}, we quantify the acceleration of learning with the number of epochs needed to reach 5\% of the initial loss. We observe that convergence time smoothly decreases as a function of connectivity strength g; for very large g, networks finally transition to chaotic activity \cite{sompolinsky1988chaos}, and convergence time increases again. 

After observing the drastic decrease in learning time, we wondered how initial connectivity affects the resulting connectivity changes. 
The first observation is that, for increasing $g$, the final connectivity $W = W_0 + \Delta W$ is dominated by $W_0$, since $||W_0|| = \sqrt{N}g$. In fact, the norm of $\Delta W$ not only remains unchanged for increasing $N$ (see supplementary), but further decreases with increasing $g$, see \cref{fig:learning_time_min_rank}\textbf{(b)}. If a smaller $\Delta W$ solves the task for larger initial connectivity, it is reasonable to assume that $W_0$ amplifies the effect of $\Delta W$. To test this idea, we shuffled the elements of $W_0$, destroying any correlation between $W_0$ and $\Delta W$, while maintaining its statistics. 
The loss after replacing the connectivity with $W_0^\mathrm{shuffle} + \Delta W$ is shown in \autoref{fig:learning_time_min_rank}\textbf{(c)}. For all tasks, shuffling strongly degraded performance except for cases with very weak initial connectivity. 

\paragraph{Low-rank changes in connectivity}
Despite the effects of the initial connectivity on convergence time and the norm of $\Delta W$, the low-rank nature of $\Delta W $ remains similar to the case with $g=0$. In \cref{fig:all_tasks}\textbf{(g-h)}, the SVs of $\Delta W$ are plotted in lighter colors. We see that the pattern and overall amplitude is very similar to the darker lines for $g=0$: only a small number of SVs dominates over a tail. To assess the functional rank, we replaced $\Delta W$ in our RNN with the rank-$R$ truncation, \cref{eq:dw_trunc}, while keeping the initial connectivity $W_0$ identical. The resulting loss, \cref{fig:all_tasks}\textbf{(j-l)}, indicates that the effective connectivity change is indeed low-rank: for all three tasks, it drops to a value close to zero before rank 10. We quantified this observation by computing the ``functional rank'', the rank at which the loss decreases below 5\% of the initial value [see \cref{fig:learning_time_min_rank}\textbf{(d)}]. This functional rank is between 2 and 10 for all three tasks (averaged over independent simulations). It increases with $g$ for the flip-flop task, while it remains less affected for the other two tasks.

\section{Analytical results for linear system}
The observation of effective low-rank changes in connectivity and accelerated learning for random initial connectivity were general across the three different tasks considered. To understand the underlying mechanisms, we turn to a much simpler task and a linear RNN model. This setting allows us to analytically describe the learning dynamics, understand the origin of the low-rank connectivity changes, and quantify how correlations between $W_0$ and $\Delta W$ accelerate learning. Our approach is similar to that of \citet{saxe2019mathematical}, who analyzed gradient descent dynamics in linear feed-forward networks. Both for the feed-forward and the recurrent model, the learning dynamics are nonlinear despite the linearity of the networks. Nevertheless, we will see that the recurrent nature of our models results in very different dynamics compared to the linear feed-forward model. Below we will present our main results for the simplified model; the details of all our analytical derivations can be found in the supplementary. 

\begin{figure}
  \centering
  \includegraphics[width=\figwidth]{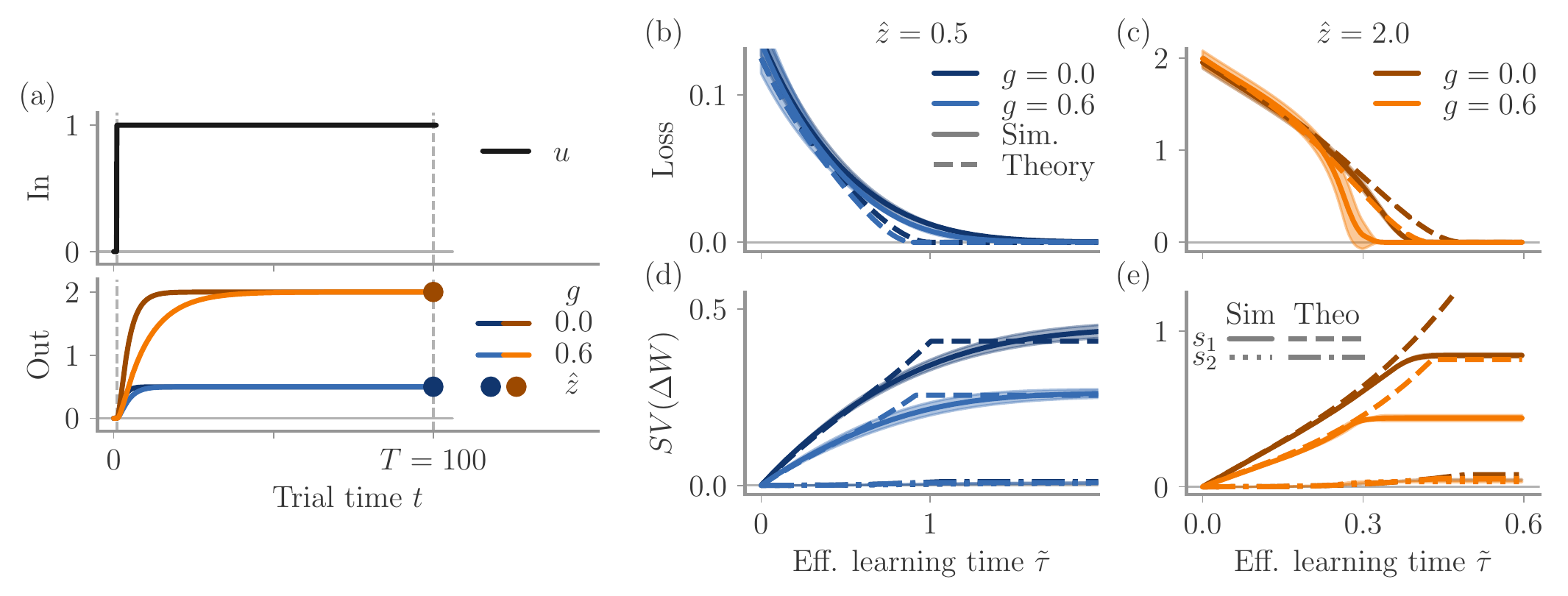}
  \caption{
  Learning a simple input-output transformation in a linear network. 
  \textbf{(a)} Task summary. Output for trained networks with two different 
  initial connectivity strengths $g = 0.0,\, 0.6$ and target amplitudes $\hat{z} = 0.5,\, 2.0$. 
  Input starts at $t=1$, loss is evaluated at $T = 100$. 
  \textbf{(b,c)} Loss over training for target values $\hat{z}=0.5$ and $\hat{z} = 2.0$. Full lines indicate simulation results, dashed lines our theoretical prediction. 
  \textbf{(d,e)} First two SVs of $\Delta W$ at the end of training (full, dotted lines) and theoretical predictions (dashed, dashed-dotted). 
  In panels \textbf{(b-e)}, the simulation results are averaged over five independent instances. Shades, if visible, indicate the standard deviation. 
  Note that the x-axes in \textbf{(b-e)} show the rescaled, effective learning time $\tilde{\tau} = \beta^2 \tau$, with $\beta = 1/(1 - g^2)$.
  Simulation parameters: $N = 1024$, training for 200 epochs with learning rate $\eta$ adapted (see supplementary).
  }
  \label{fig:linear_explain_loss_sv}
\end{figure}

\paragraph{Simplified setting}
Our simple task is an input-output transformation: Given a constant input $u(t) = 1$, the output $z(t)$ has to reach a target value $\hat{z}$ at time $T$. The corresponding loss is $L = (\hat{z} - z(T))^2 / 2$. An example with two different target values $\hat{z} = 0.5,\, 2.0$ is plotted in \cref{fig:linear_explain_loss_sv}\textbf{(a)}. 
The linear RNN model is obtained by replacing the nonlinearity in \cref{eq:dxdt_nonlin} with the identity, $\phi(\mathbf{x}) =\mathbf{x}$, and keeping only a single input and output. All weights are initialized as before. We keep the initial connectivity strength $g < 1$ so that the linear network remains stable. To further simplify, we constrain weight changes to the recurrent weights $W$ only, and apply plain gradient descent. To compare between different simulations, we define the learning time $\tau = \eta \cdot \text{epochs}$.

Evaluating the trained networks reveals similar phenomena as observed for the nonlinear, more complex tasks. \Cref{fig:linear_explain_loss_sv}\textbf{(b-e)} shows the loss and SVs of $\Delta W$ over learning time for two values of $g$. We observe that learning induces low-rank connectivity changes -- in fact, a single SV dominates. Because of the small magnitude of the second SV, truncating $\Delta W$ at rank 1 does not lead to increased loss (not shown), so that the functional rank as defined in the previous section is 1. Comparing between $g=0$ and $g=0.6$, we further see that learning is accelerated by the initial connectivity, and that the magnitude of the first SV decreases with increasing $g$. These observations will be quantified with our analytical results.

\paragraph{Gradient descent dynamics}
For our analytical treatment, we only consider the limit of long trials, with the output $z = \lim_{T\to\infty} z(T)$ at the end of a trial. In this limit, the network converges to its fixed point $\mathbf{x}^*=\sqrt{N} \left( I - W \right)^{-1} \mathbf{m}$ with identity matrix $I$, and the readout is
\begin{equation} 
    \label{eq:fp}
    z 
    = \frac{\mathbf{w}^T\! \mathbf{x}^*}{\sqrt{N}}
    = \mathbf{w}^T\! \left( I - W \right)^{-1} \mathbf{m}  \,.
\end{equation}
The input and output vectors, $\mathbf{m}$ and $\mathbf{w}$, remain fixed during training, and only $W$ is changed. We can explicitly compute the changes induced by the gradient of the loss:
\begin{equation}
    \label{eq:grad_desc_full}
    \frac{\mathrm{d} W(\tau) }{\mathrm{d} \tau}
    = -\frac{\mathrm{d}L}{\mathrm{d}W}
    = \left[\hat{z} - z(\tau) \right] 
    \, \left[ I - W^T\!(\tau) \right]^{-1}
    \mathbf{w} 
    \mathbf{m}^T
    \left[ I - W^T\!(\tau) \right]^{-1}  \,,
\end{equation}
with initial connectivity $W(0) = W_0$. We made a continuous-time approximation of the weight updates (``gradient flow''), valid to small learning rates $\eta$. Note that the readout $z$ at the fixed point depends on the learning time $\tau$ through $W(\tau)$. 

Note that, unlike the feed-forward case \cite{saxe2013exact}, the inverse of $W$ appears in \cref{eq:grad_desc_full}, opening the possibility of divergence during learning.
It also precludes a closed-form solution to the dynamics.
However, we can obtain analytical insight by expanding the learning dynamics in learning time around the initial connectivity \cite{bender2013advanced}. We write
\begin{equation}
    W(\tau) = \sum_{k=0}^{\infty} W_k \,\frac{\tau^k}{k!}  \,.
\end{equation}
The changes in connectivity are obtained by subtracting $W_0$, which yields $\Delta W(\tau) = W_1 \tau + W_2 \tau^2 / 2 + \dots$. We analytically computed the coefficients $W_k$ by evaluating $\mathrm{d}^k W/\mathrm{d}\tau^k$ at $\tau = 0$. A comparison of the expansion up to third order with the numerical results from gradient descent learning indicates close agreement during most of the learning [see \cref{fig:linear_explain_loss_sv}\textbf{(b-e)} full vs. dashed lines].

\paragraph{Learning dynamics in absence of initial connectivity}
It is instructive to first consider the case of no initial connectivity, $g=0$. The readout at the beginning of training is then $z_0 = \mathbf{w}^T\!\mathbf{m}$. Due to the independence of $\mathbf{m}$ and $\mathbf{w}$, the expected value of $z_0$ vanishes. Moreover, the standard deviation scales as $1 / \sqrt{N}$ with the network size. In this work, we are interested in the learning dynamics for large networks; all our analytical results are valid in the limit $N \to \infty$. We therefore write $z_0 = 0$. Similar reasoning goes for all scalar quantities of interest: they are of order $\mathcal{O}(1)$, with deviations $\mathcal{O}(1/\sqrt{N})$. With this self-averaging quality, we omit stating the limit as well as the expectation symbol and use the equality sign instead. 

Inserting $W_0$ and $z_0$ -- both zero -- into the gradient descent, \cref{eq:grad_desc_full}, yields the first order coefficient
\begin{equation}
    W_1 = \hat{z} \mathbf{w} \mathbf{m}^T\! \,.
\end{equation}
Hence, the weight changes at linear order in $\tau$ are described by a rank-one matrix, and the readout is 
$z(\tau) = \tau \hat{z} + \mathcal{O}(\tau^2)$. The gradient descent for $g=0$ would therefore converge at $\tau_1^*=1$, if it only depended on the first-order term.
The numerical results already show deviations in the form of faster or slower convergence, depending on the target $\hat{z}$ [see dark lines in \cref{fig:linear_explain_loss_sv}\textbf{(b,c)} and note that $\tilde{\tau} = \tau$ for $g=0$]. This indicates the importance of higher order terms.

We observe that the gradient in \cref{eq:grad_desc_full} contains the transpose $W^T$. At higher orders, this term introduces other outer-product combinations of $\mathbf{m}$ and $\mathbf{w}$. In fact, for $g=0$, these are the only vectors present in the gradient, so that the connectivity can always be written as 
\begin{equation}
    \label{eq:W_g0}
    \Delta W(\tau) = 
    \begin{bmatrix}
    \mathbf{w} & \mathbf{m}
    \end{bmatrix}
    \begin{bmatrix}
    A_{11} & A_{12} \\ A_{21} & A_{22}
    \end{bmatrix}
    \begin{bmatrix}
    \mathbf{w}^T \\ \mathbf{m}^T
    \end{bmatrix} \,.
\end{equation}
This form implies that $\Delta W$ will be at most a rank-two matrix. An analysis of the SVs [\cref{eq:svs} below for general $g$] reveals that the second SV remains very small, as visible in \cref{fig:linear_explain_loss_sv}\textbf{(d,e)}.

The entries of the $2 \times 2$ matrix $A(\tau)$ up to order $\mathcal{O}(\tau^3)$ are (see supplementary)
\begin{equation}
    A_{11}
    = \frac{\hat{z}^2}{2} \left(\tau^2  - \tau^3\right) \,,
    \qquad 
    A_{12} = \hat{z} \left(\tau - \frac{\tau^2}{2} + \frac{\tau^3}{6} (1 + 2 \hat{z}^2) \right) \,,
    \qquad
    A_{21} = \frac{\hat{z}^3\tau^3}{3} \,,
\end{equation}
and $A_{22} = A_{11}$. The first surprising observation is that the target value $\hat{z}$ enters nonlinearly into the expressions above. This is the origin of the qualitative difference between learning curves for different values of the target output in \cref{fig:linear_explain_loss_sv}\textbf{(b,c)}. 

We further observe that the connectivity changes develop a nonzero eigenvalue only at $\mathcal{O}(\tau^2)$. This is because the off-diagonal terms, which grow linearly with $\tau$ contribute a zero eigenvalue because $\mathbf{m}^T\!\mathbf{w} = 0$. At second order the diagonal entries of $A$ -- and, with it, the eigenvalues -- change. Changes in connectivity eigenvalues imply changes in time scales of network dynamics, which may be necessary for some tasks (for example, those involving memory), but can also lead to problems of exploding gradients (see supplementary). 

\begin{figure}
  \centering    
  \includegraphics[width=\figwidth]{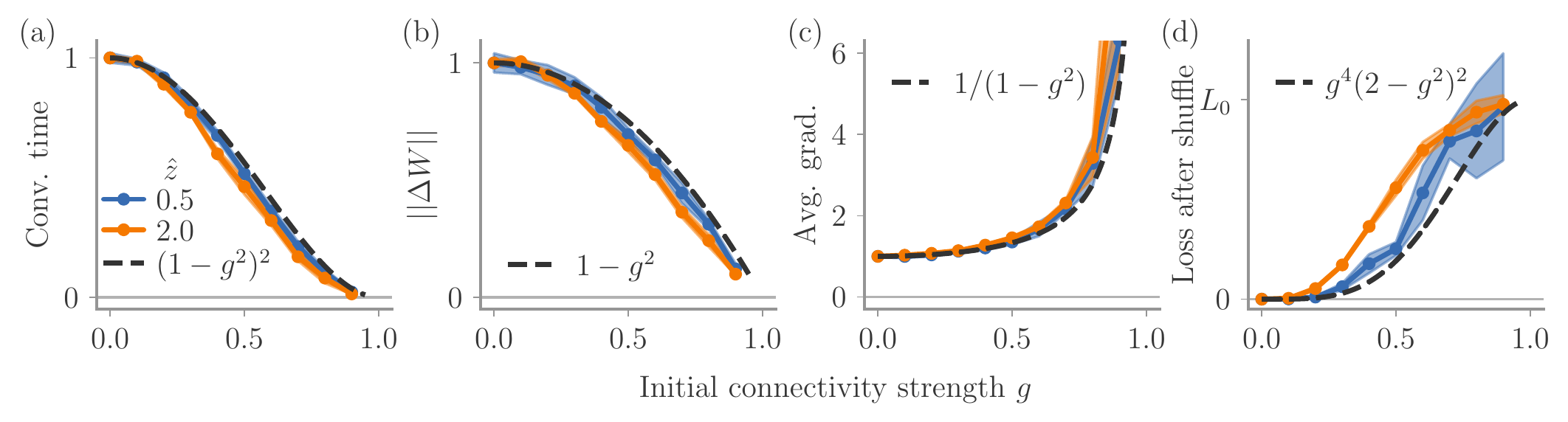}
  \caption{
  Dependence of learning dynamics on initial connectivity strength $g$ in the simplified task.
  \textbf{(a)} Learning time $\tau^*$ until loss reached 5\% of its initial value. 
  \textbf{(b)} Norm of final weight changes $\Delta W$. 
  \textbf{(c)} Norm of gradient $\mathrm{d}W /\mathrm{d}\tau$, averaged over the interval $\tau \in [0, \tau^*]$.
  The quantities in panels \textbf{(a-c)} are normalized by their value at $g=0$. 
  \textbf{(d)} Loss after shuffling the initial connectivity $W_0$, normalized by initial loss.
  In all panels, thick full lines indicate the average over five independent simulations, shades the standard deviation,
  and dashed lines the first-order theoretical prediction. 
  }
  \label{fig:linear_lt}
\end{figure}

\paragraph{Effects of initial connectivity}
In the presence of initial connectivity, we can still apply the expansion introduced above. Due to the independence of $W_0$, $\mathbf{m}$, and $\mathbf{w}$, the initial readout $z_0$ remains zero. The gradient descent, \cref{eq:grad_desc_full}, then directly yields the first-order connectivity coefficient 
\begin{equation}
    \label{eq:W_1g}
    W_1 = \hat{z} \, B^T\! \mathbf{w} \, \mathbf{m}^T\! B^T\! 
    \,, \qquad \text{with}  \qquad
    B = (I - W_0)^{-1} \,.
\end{equation}
Thus, $W_1$ is still a rank-one matrix despite the full-rank initial connectivity.
However, the connectivity changes now include the initial connectivity $W_0$ via the matrix $B$.
As a consequence, the norm of the first-order coefficient, $|| W_1 || = \hat{z} \beta$ (see supplementary), increases with $g$ by the factor 
\begin{equation}
    \beta 
    =\mathbf{w}^T\! B B^T\! \mathbf{w}
    =\mathbf{m}^T\! B^T\! B \mathbf{m}
    = \frac{1}{1 - g^2} \,.
\end{equation}
The readout is also affected by the initial connectivity. We compute (see supplementary)
\begin{equation}
    z(\tau) = \tau \hat{z} \beta^2 + \mathcal{O}(\tau^2) \,.
\end{equation}
Learning converges when $z(\tau)$ reaches the target value $\hat{z}$. The first-order prediction of the 
convergence time is therefore $\tau_1^* = 1 / \beta^2$, and the initial connectivity accelerates learning by the factor $1/\beta^2 = (1 - g^2)^2$. 
We can decompose this acceleration into two factors: 
The growth rate is increased by $\beta$, and the norm of the final connectivity changes decreased by $1/\beta$. For the first contribution, we note that the first-order coefficient $W_1$ is, by definition, the constant part of the gradient, and hence the rate at which connectivity changes. For the second contribution, we compute the norm of $\Delta W(\tau)$ at the predicted convergence time $\tau^*_1$ (see supplementary). 

In \cref{fig:linear_lt}\textbf{(a-c)}, we compare our first-order predictions with numerical simulations. In panels \textbf{(a,b)}, we plot the convergence time $\tau^*$ and the norm of $\Delta W$ at the end of training. As for the more complex, nonlinear tasks [see \cref{fig:learning_time_min_rank}\textbf{(a,b)}], we defined the numerical $\tau^*$ as the point in time where the loss drops to 5\% of the initial value. For the gradient, panel \textbf{(c)}, we averaged the norm $||\mathrm{d}W / \mathrm{d}\tau||$ over the interval $[0, \tau^*]$. To compare the collapsed curves with the predicted scalings, we normalized the curves for the different target values $\hat{z}$ by their value at $g=0$ for all three quantities. We observe good agreement between the numerical results and the theory, even though we only used the first-order predictions, and $\tau^*$ often shows notable differences between theory and simulation [for example in \cref{fig:linear_explain_loss_sv}\textbf{(b,c})]. 

Finally, we assess the role of correlations between $\Delta W$ and $W_0$ by shuffling $W_0$. After shuffling, the readout loses the amplification by $\beta^2$ and is hence $z^\mathrm{shuff} = \tau^*_1 \hat{z}$. The corresponding loss is $L_1^\mathrm{shuff} = L_0\, g^4 (2 - g^2)^2$, with initial loss $L_0 = \hat{z}^2/2$. A comparison of this first-order prediction with numerical results shows qualitative agreement with notable quantitative differences especially for the larger target amplitude, see \cref{fig:linear_lt}\textbf{(d)}. A comparison with the nonlinear case, \cref{fig:learning_time_min_rank}\textbf{(c)} shows that our simple model captures the phenomenon qualitatively.

\paragraph{Higher-order terms}
Does the initial connectivity lead to higher-rank changes in connectivity? For $g>0$, the explicit rank-two expression for the weight changes, \cref{eq:W_g0}, does not hold anymore: The input and output vectors accumulate multiples of $B$ and $B^T$ (such as $B^T\!\mathbf{\mathbf{w}}$ and  $B B^T\! \mathbf{w}$) which increase the number of possible outer products -- and hence potentially the rank. However, computing the first two SVs, $s_1$ and $s_2$, up to order $\mathcal{O}(\tau^3)$ (see supplementary) shows that $\Delta W$ remains approximately rank one: 
\begin{equation}
    \label{eq:svs}
    s_1 = 
        \frac{\hat{z}}{\beta}
        \left[
        \tilde{\tau}
        - \frac{\tilde{\tau}^2}{2}  
        + \left(1 + \frac{7}{2} \hat{z}^2 \beta
        \right) \frac{\tilde{\tau}^3}{6}  \right]\,, \qquad 
    s_2 = \hat{z}^3\frac{\tilde{\tau}^3}{12} \,.
\end{equation}
where $\tilde{\tau} = \beta^2 \tau$ is the effective learning time. We observe that $s_1$ grows linearly, but $s_2$ only at third order of $\tau$. Different parts of connectivity therefore grow on top of each other, giving rise to a temporal hierarchy in the learning dynamics.
Numerical simulations show good agreement with this prediction (see supplementary).

We further state the resulting readout up to $\mathcal{O}(\tau^3)$:
\begin{equation}
    \label{eq:z}
    z(\tau)
    = \hat{z} 
    \left[
    \tilde{\tau}
    -  \frac{\tilde{\tau}^2}{2}
    + (1 + 8 \hat{z}^2 \beta) \frac{\tilde{\tau}^3}{6}
    \right] \,.
\end{equation}
The appearance of $\beta$ in the third-order contributions in \cref{eq:svs,eq:z} shows that the learning with different values of $g$ does not entirely collapse onto one curve after rescaling the time by $\beta^2$. Instead, there is an additional acceleration, which increases with increasing target amplitude $\hat{z}$. This effect can be appreciated in \cref{fig:linear_explain_loss_sv}\textbf{(b,c)}, where for larger $\hat{z}$ the loss curve becomes concave. Note that our approximation up to $\mathcal{O}(\tau^3)$ predicts this trend, despite quantitative disagreement. As we saw in \cref{fig:linear_lt}, the scaling of the convergence time $\tau^*$ with $g$ is not strongly affected by the higher order terms.

\section{Beyond neuroscience tasks}
We asked whether our observation that connectivity changes are low-rank despite full-rank initial connectivity would extend to more complex network architectures and tasks, specifically those not restricted to a small input or output dimension. We therefore trained a two-layer LSTM network on a natural language processing task, sentiment analysis of movie reviews \cite{socher2013recursive} (details in supplementary). 

The SVs at the end of training showed the pattern that we predicted: learning only leads to small changes in the connectivity so that the final connectivity $W$ is dominated by the initial connectivity and has full rank.  The changes $\Delta W$ only have a small number of large SVs. For the recurrent weights of layer 2, the SVs are plotted in \cref{fig:nlp}\textbf{(a)}; other weights behave similarly (see supplementary).

Like before, we evaluated the accuracy of networks after truncation at a given rank, see \cref{fig:nlp}\textbf{(b)}. We truncated the recurrent weights of both layers as well as input weights to layer 2. 
If we keep the random parts and  truncate the changes as in \cref{eq:dw_trunc} a rank-10 approximation already yields the final training accuracy. In contrast, if we truncate the entire weight matrices, as previously suggested \cite{winata2019effectiveness}, it takes more that half of the network rank (256 neurons per layer) to get close to the final accuracy. 


\begin{figure}[tb]
\centering
\includegraphics[width=1.0\textwidth]{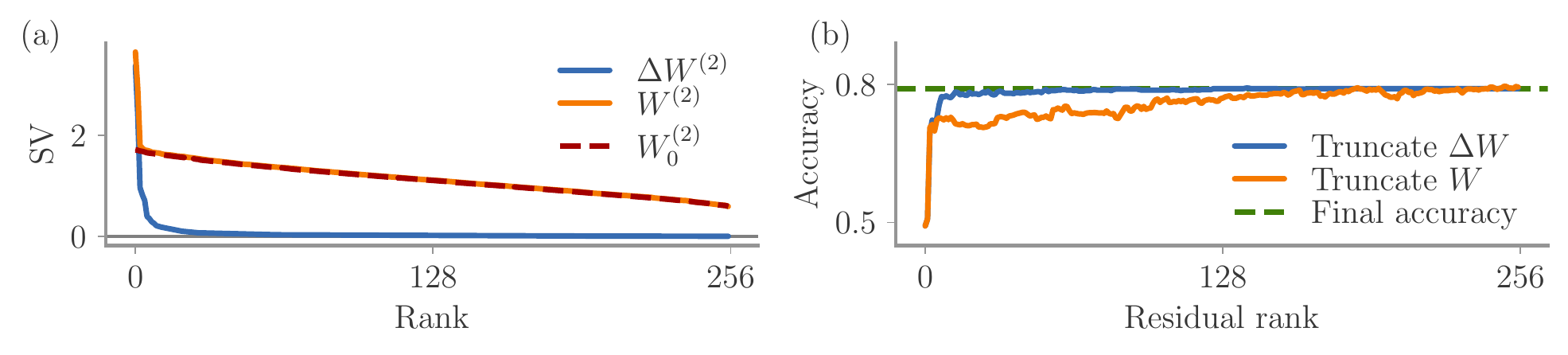}
\caption{
    Low-rank changes for a two-layer LSTM model trained on a sentiment analysis task.
    \textbf{(a)} Singular values (SVs) of the recurrent weights in the second layer (256 neurons). The initial, random $W_0$ is full rank, and the final $W$  visibly differs from it only for the first SVs. The changes, $\Delta W$, are approximately low-rank. 
    \textbf{(b)} Validation accuracy after truncating the lower singular values of connectivity. We either truncated $W$ directly, or applied truncation only to $\Delta W$ while keeping $W_0$. We truncated the recurrent weights of both layers and the input weights of layer 2. 
}
\label{fig:nlp}
\end{figure}

\section{Discussion}

\paragraph{Summary of results}
Our key finding is that the connectivity changes $\Delta W$ induced by unconstrained training on low-dimensional tasks are of low rank.
With our simplified analytical model, we demonstrated why: The connectivity changes are spanned by a small number of existing directions, determined by the input and output vectors. Without initial connectivity, the maximum rank that linear networks can obtain through learning is in fact bounded by this number. The initial connectivity $W_0$ enlarges the pool of available directions. The fact that learning arrives at a low-rank solution even in presence of initial connectivity is then a result of the temporal structure of learning: Initially, only a small number of available directions grow, inducing a low-rank structure. For our simplified task, the first of these structures already reduces the loss, and learning converges before other structures emerge; the final connectivity changes are hence rank-one. For other tasks, the available input and output directions alone may not be sufficient, so that initial connectivity becomes necessary for successful learning (see supplementary). Note that our theoretical analysis is limited to linear networks; however, nonlinearity may also contribute to generate novel learning directions. 

Our numerical simulations further showed that initial connectivity significantly accelerated learning. Our analytical results revealed the underlying mechanism: The input and output vectors spanning the gradient are multiplied by powers of $W_0$, which strongly correlates $\Delta W$ to $W_0$. This correlation amplifies the effect of $\Delta W$, and removing the correlation by shuffling $W_0$ indeed degrades performance. This is in line with a recent study demonstrating such amplification through correlation between a random matrix and a low-rank perturbation in a model without learning \cite{schuessler2020dynamics}.

Finally, we showed that the general observation of low-rank weight changes indeed holds even in a much more complex setting: a sentiment analysis task and a two-layer LSTM network. This implies a large potential for network compression \cite{winata2019effectiveness}: one may truncate the changes in connectivity at a very low rank and recover the specific random initial connectivity using the seed of its random number generator.

\paragraph{Task dimension and rank}
Low-rank connectivity structures have previously been studied and applied. On the one hand, a number of RNN frameworks explicitly rely on low-rank feedback for training \cite{jaeger2004harnessing,eliasmith2004neural,sussillo2009generating,logiaco2019model,barak2020mapping}. On the other hand, low-rank networks are amenable to analysis, because the network activity is low-dimensional and evolves in directions determined by the vectors spanning the connectivity \cite{hopfield1982neural,tirozzi1991chaos,mastrogiuseppe2018linking,rivkind2017local,schuessler2020dynamics}. Our surprising observation that unconstrained gradient descent also leads to low-rank connectivity opens new possibilities for studying general gradient-based learning with the tools developed by previous works. 

We observed that the functional rank of the training-induced connectivity changes is strongly task dependent. A better understanding of the relation between task and connectivity calls for a concept of a task dimension, ideally based on the underlying abstract computations and independent of the specific implementation \cite{gao2017theory,yang2019task,maheswaranathan2019universality,li2018measuring}. Such a concept would allow to compare the solutions obtained by different algorithms and define a necessary minimal rank for a given task \cite{dubreuil2020complementary}. 

\paragraph{Learning as a dynamical process and relation to feed-forward networks}
Our approach stresses a dynamical perspective on learning, in which the solutions are not determined by the task alone, but also by the initial connectivity and the temporal evolution of weight changes. In particular, our expansion in learning time shows that some components in the connectivity only grow after others are present, which induces a temporal hierarchy. This affects the solutions the network arrives at. The temporal structure may also induce pitfalls for learning, for example divergent gradients when the networks undergo a phase transition \cite{pascanu2013difficulty} (see supplementary). A better understanding of the learning dynamics could be used to circumvent such problems, for example by introducing adapted learning curricula \cite{bengio2009curriculum}. 

Learning in feed-forward networks has previously been analyzed from a similar perspective. It was found that the statistical structure of the training data induces a temporal hierarchy with long plateaus between step-like transitions in the learning curve \cite{saxe2013exact,saxe2019mathematical,advani2017high,lampinen2018analytic,yoshida2019data,goldt2019modelling}. The hierarchy in our work originates in the dynamics of the RNN rather than the structure of the training data. For example, the plateaus seen in \cref{fig:all_tasks}\textbf{(d-f)} can be related to phase transitions in the network dynamics, such as the emergence of new fixed points. Combining such internal learning dynamics with structured training data would be an interesting future direction. 


Finally, recent work on feed-forward networks identified two different learning regimes: a kernel regime vs. a rich, feature-learning regime \cite{jacot2018neural,chizat2018global,arora2019fine,woodworth2020kernel}. In the prior, the change in weights vanishes as the network width increases, and the network function can be linearized around the weights at initialization. In our work, too, the weight changes $\Delta W$ become infinitely small in the limit of wide networks. However, even such vanishing $\Delta W$ may significantly change the dynamics of the neural network by inducing large outlier eigenvalues \cite{schuessler2020dynamics}. For example, the readout for our linear network, \cref{eq:fp}, diverges for a eigenvalue of $W$ approaching 1. In such a case, the network function cannot be approximated by linearization around the initial weights. Understanding the relation between learning regimes in feed-forward and recurrent networks constitutes an interesting field for future studies.

\newpage
\section*{Broader Impact}

This work is a theoretical study on the dynamics of learning in RNNs. We show which kind of connectivity changes are induced by gradient descent. We expect that our insights will help to understand learning in RNNs, which benefits the research community as a whole and may ultimately lead to the development of improved learning algorithms or schemes. As a possible application, we show that one can use our results to efficiently compress a multi-layer RNN trained on a natural language processing task. 
In this work, there are no new algorithms, tasks, or data sets introduced. Therefore, the questions regarding any disadvantages, failures of the system, or biases do not apply.

\begin{ack}
This work was supported in part by the Israeli Science Foundation (grant number 346/16, OB).
The project was further supported by the  ANR project MORSE (ANR-16-CE37-0016), the program “Ecoles Universitaires de Recherche” launched by the French Government and implemented by the ANR, with the reference ANR-17-EURE-0017.
F.S. acknowledges the Max Planck Society for a Minerva Fellowship.
There are no competing interests.
\end{ack}

{\small
\pagestyle{empty}
\bibliography{neurips_2020_arxiv}
}

\newpage
\beginsupplement
\section*{\LARGE{Supplementary information}}
\section{Simulation parameters}

All simulations were based on pytorch \citeSM{pytorch}. For the nonlinear neuroscience tasks, we applied the gradient descent method ``Adam'' \citeSM{kingma2014adam} to the recurrent weights $W$ as well as to the input and output vectors $\mathbf{m}_i$, $\mathbf{w}_i$. We checked that our results did not depend qualitatively on the choice of the ``Adam'' algorithm over plain gradient descent; however, training converged more easily for this choice of algorithm. We also checked that restricting training to $W$ only (as for the simple model) did not alter our results qualitatively (although, with this restriction, training on the Romo task for small values of $g$ did not converge). Code for reproducing our results can be found on \url{https://github.com/frschu/neurips_2020_interplay_randomness_structure/}.

The network size for the results in Figures 1 and 2 was $N = 256$, and the learning rate $\eta = 0.05 / N$. We trained the networks for a maximum number of 1000, 2000, and 6000 epochs for the flip-flop, Mante, and Romo task, respectively. Each epoch consisted of a batch of 32 independent task trials. For evaluation of the loss after rank-truncation or shuffling $W_0$, we used a single batch of 512 independent task trials. 
Note that for ``Adam'', the learning rate is scaled with $N$ to obtain approximate invariance of the loss curve for different network sizes $N$. 
Further note that Fig 1 does not always show the loss over all learning epochs (so that the differences in the initial phase are more clearly visible). 

For the simpler, linear model, we applied plain gradient descent and only adapted $W$. We trained all models for $200$ epochs, and the learning rate was adapted in order to obtain smooth convergence within these 200 epochs. We set $\eta = \eta_0 (1 - g^2)^2$, with $\eta_0 = 0.015,\, 0.003$ for $\hat{z} = 0.5,\, 2.0$, respectively. We checked that our numerical results do not depend on this choice, as long as a sufficiently small learning rate and large enough number of epochs is chosen. 

The network dynamics are described by the continuous dynamics
\begin{equation}
    \label{eq:dxdt_nonlin_supp}
    \dot{\mathbf{x}}(t) = -\mathbf{x}(t) + W \phi(\mathbf{x}(t)) + \sqrt{N} \sum_{i=1}^{N_\mathrm{in}} \mathbf{m}_i u_i(t) \,,
\end{equation}
with initial condition $\mathrm{x}(0) = \bm{0}$. For the simulation, we discretized these using the Euler-forward scheme:
\begin{equation}
    \label{eq:dxdt_nonlin_discr_supp}
    \mathbf{x}_{k+1} = 
        (1 - \Delta t)\mathbf{x}_k 
        + \Delta t \left[W \phi(\mathbf{x}_k) + \sqrt{N} \sum_{i=1}^{N_\mathrm{in}} \mathbf{m}_i u_{i,k} \right]\,,
\end{equation}
with a discrete time step $\Delta t = 0.5$ and $\mathbf{x}(t = k \Delta t) = \mathbf{x}_{k}$. We checked that our results did not change qualitatively for choosing a smaller $\Delta t$ or fully discrete dynamics ($\Delta t = 1$). 

For the gradient-based updates, we defined the quadratic loss 
\begin{equation}
    l(t) = 
    \frac{1}{N_\mathrm{out}} \sum_{i=1}^{N_\mathrm{out}} \frac{1}{2} \left[z_i(t) - \hat{z}_i(t)\right]^2 \,,
\end{equation}
with readout $z_i(t)$, target $\hat{z}_i(t)$, and number of outputs $N_\mathrm{out}$.
Depending on the task, the loss was defined only during specific times of the task (during decision or fixation periods, see task descriptions).
Accordingly, for each task we defined a boolean mask $M_k$, indicating the points $k$ on the discrete time grid were the loss was active. The full loss was the average over this mask: 
\begin{equation}
    L = \frac{1}{N_M} \sum_{k=1}^{k_\mathrm{max}} M_k \, l (k \Delta t) \,,
\end{equation}
with $N_M = \sum_{k=0}^{k_\mathrm{max}} M_k$, $k_\mathrm{max} = T / \Delta t$ and trial time $T$. 

\section{Task details}
\begin{table}[htb]
    \centering
    \caption{Task parameters}
    \begin{tabular}{llrrrr}  
    \toprule
Parameter           & Symbol            & Flip-flop     & Mante     & Romo      & Simple task\\
\midrule
\# inputs           & $N_\mathrm{in}$   & 2             & 4         & 1         & 1 \\
\# outputs          & $N_\mathrm{out}$  & 2             & 1         & 2         & 1 \\
Trial duration      & $T$               & 50            & 48        & 30        & 101 \\
Fixation duration   & $t_\mathrm{fix}$  & 1             & 3         & 3         & 1 \\
Stimulus duration   & $t_\mathrm{stim}$ & 1             & 20        & 1         & - \\
Decision delay      & $t_\mathrm{delay}$& 5             & 5         & 5         & - \\ 
Stimulus delay      & $t_\mathrm{sd}$   & $\mathcal{U}(5, 25)$ & - & $\mathcal{U}(2, 8)$ & - \\
Decision duration   & $t_\mathrm{dec}$  & -             & 20        & 10        & 1 \\
Input amplitude     & $u_\mathrm{amp}$  & 1             & 1         & $\mathcal{U}(0.5, 1.5)$ & 1 \\
Target amplitude    & $\hat{z}_\mathrm{amp}$  & 0.5     & 0.5       & 0.5       & $\{0.5, 2.0\}$  \\
    \bottomrule
    \end{tabular}
    \label{tab:task_parameters}
\end{table}

All task share a broad overall structure: a trial of length $T$ contains an initial ``fixation'' period without input of length $t_\mathrm{fix}$, followed by the first input. During each input phase of duration $t_\mathrm{stim}$, all or some of the inputs $u_i$ have a nonzero value with amplitude $u_\mathrm{amp}$. Finally, there are distinct decision periods during which the target $\hat{z}$ is nonzero, with amplitude $\hat{z}_\mathrm{amp}$. The decision periods are preceded by a decision delay, in which the loss is inactive, and which allows the output to converge to the target value. For the flip-flop task and the simple task, the loss is inactive outside of the decision periods; for the Mante and Romo tasks, all output channels are supposed to stay at zero until the beginning of the decision delay (the corresponding target is $\hat{z}_i = 0$ for all channels $i$). Below, we describe further details for each task. The parameters and their numerical values used in the simulations reported in the main text are summarized in \cref{tab:task_parameters}. 

\paragraph{Flip-flop task}
During each trial, the network receives a number of short pulses of duration $t_\mathrm{stim}$. During such a pulse, one input channel is set to $u_i(t) = s \, u_\mathrm{amp}$, the others remain zero. The channel and sign $s \in \{\pm 1\}$ are chosen at random. After each pulse and a following delay period $t_\mathrm{delay}$, a decision period starts (the loss is activated). During the decision period, the target value is set to $\hat{z}_i(t) = s \,\hat{z}_\mathrm{amp}$. The other channel is supposed to remain silent, $\hat{z}_j(t) = 0$ for $j \ne i$. The decision period ends with the next pulse. The delays between stimuli $t_\mathrm{sd}$ are drawn randomly. Note that the plotted trial time in Fig. 1 in the main text is $T =100$, while training was done for $T = 50$. 

\paragraph{Mante task}
Each trial for the Mante task contains only a single, longer input period of duration $t_\mathrm{stim}$.
Half of the input channels correspond to the signal $u_{i}(t)$,  the other half to a context variable $u_{N_s + i}(t)$, with number of signals $N_s = N_\mathrm{in} / 2$. The signals each consist of a constant mean and random noise part: $u_{i}(t) = u_\mathrm{amp} [s_{i} + a_\mathrm{noise} \eta_{i}(t)]$ with random sign $s_{i} \in \{\pm 1\}$ and Gaussian white noise $\eta_{i}(t)$. For our simulations, we chose the relative noise amplitude $a_\mathrm{noise} = 0.05$. For the discretization, the white noise at time step $k$ is $\eta_{i, k} = n_{i, k} / \sqrt{\eta}$ with standard normal variable $n_{i, k} \sim \mathcal{N} (0, 1)$. During each trial, only a single context is active, $u_{N_s + i} = u_\mathrm{amp} \delta_{i, j}$, where $j$ is chosen randomly from the number of inputs $N_s$. Outside of the input period, all mean values of $u_i$ are set to zero (the noise terms remain active). 
The input period is followed by a decision phase of length $t_\mathrm{dec}$, with a delay $t_\mathrm{delay}$ in between. During the decision period, the output is supposed to communicate the sign $s_j$ of the relevant input $j$. The target is constant: $\hat{z}(t) = \hat{z}_\mathrm{amp} s_j$, and $\hat{z}_i(t) = 0$ for all $i \ne j$.

\paragraph{Romo task}
For the Romo task, the RNN model has only one input channel, and each trial contains two input pulses of length $t_\mathrm{stim}$ each. During the input pulses, the input is $u(t) = u_\mathrm{amp, 1}$ and $u(t) = u_\mathrm{amp, 2}$, with amplitudes drawn from a uniform distribution. Both input amplitudes are redrawn if their difference $|u_\mathrm{amp, 1} - u_\mathrm{amp, 2}|$ is below a minimal difference $u_\text{min diff} = 0.2$. The two pulses are separated by a random delay $t_\mathrm{sd}$. The end of the second pulse is followed by a delay $t_\mathrm{delay}$ and a decision period of length $t_\mathrm{dec}$. During the decision period, the output should indicate which input pulse was larger: $\hat{z}_j(t) = \hat{z}_\mathrm{amp}$ for $j = \argmax_{i}(u_\mathrm{amp, i})$. The other output should remain at zero. 

\paragraph{Simple task}
The simple task only has a single input and output channel. The input is constant starting from the end of the fixation period: $u(t) = u_\mathrm{amp}$ for $t > t_\mathrm{fix}$. The decision period is a short interval at the end of the trial, $[T - t_\mathrm{dec}, T]$. The target value during the decision period is $\hat{z}(t) = \hat{z}_\mathrm{amp}$. There is no decision delay, and the input remains constant during the decision period. Hence, this task does not contain a memory element like the other three tasks.

\section{Supplementary figures}
\FloatBarrier
\begin{figure}[hbt]
  \centering
  \includegraphics[width=\textwidth]{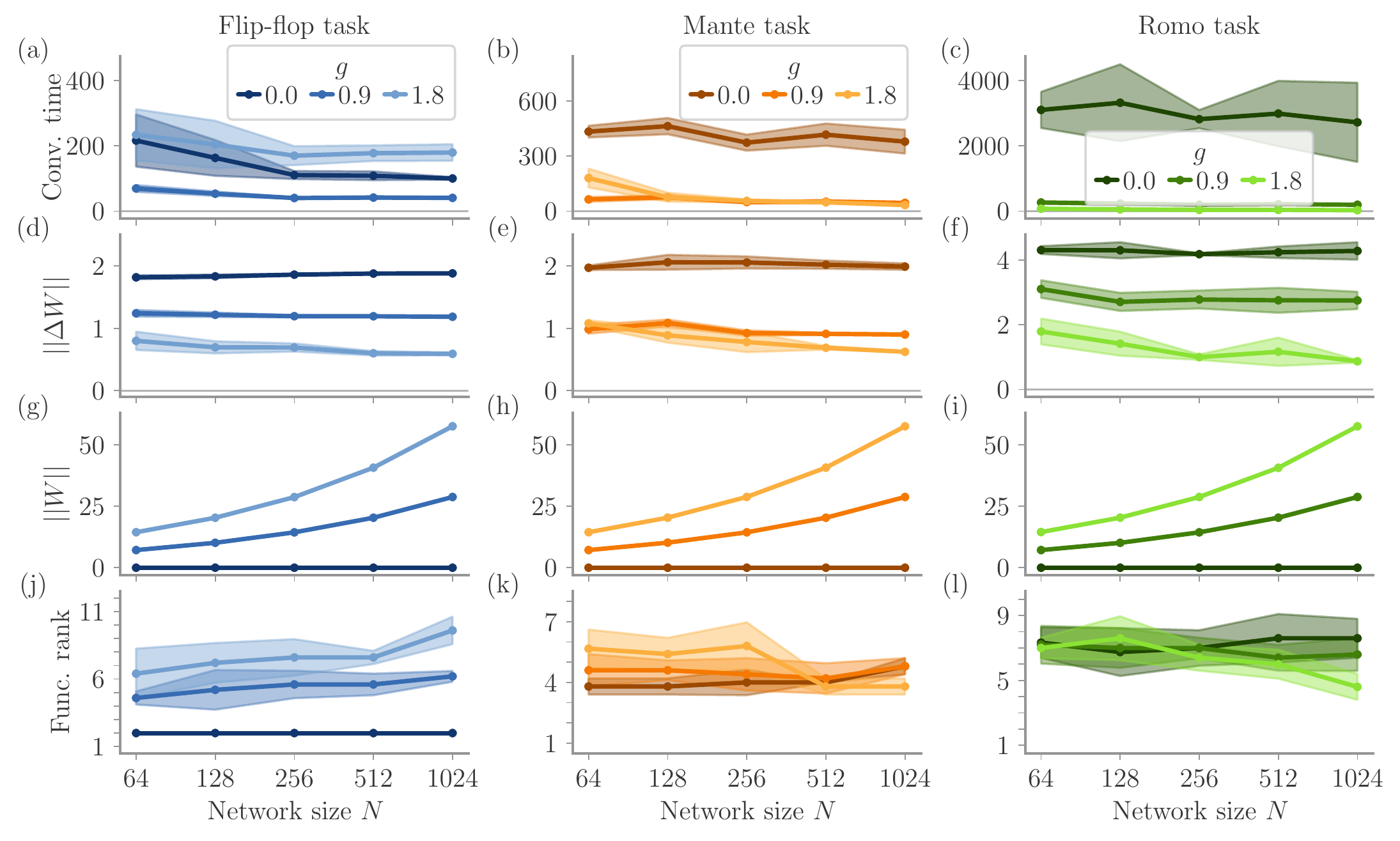}
  \caption{
  Scaling of learning dynamics with network size $N$ for all three nonlinear tasks and three different values of initial connectivity $g$ (indicated by line colors). Lines indicate average over 5 independent simulations, shades the standard deviation. Note the log-scale for networks size (x-axes). 
  \textbf{(a-c)} Number of epochs until loss reached 5\% of its initial value. 
  \textbf{(d-f)} Frobenius norm of final connectivity changes $\Delta W$. 
  \textbf{(g-i)} Frobenius norm of total connectivity $W = W_0 + \Delta W$. 
  \textbf{(j-l)} Functional rank as defined in the main text (the rank at which truncation loss falls below 5\% of the initial loss). 
  }
  \label{fig:norms_dim_recs}
\end{figure}

\begin{figure}[htb]
  \centering
  \includegraphics[width=\textwidth]{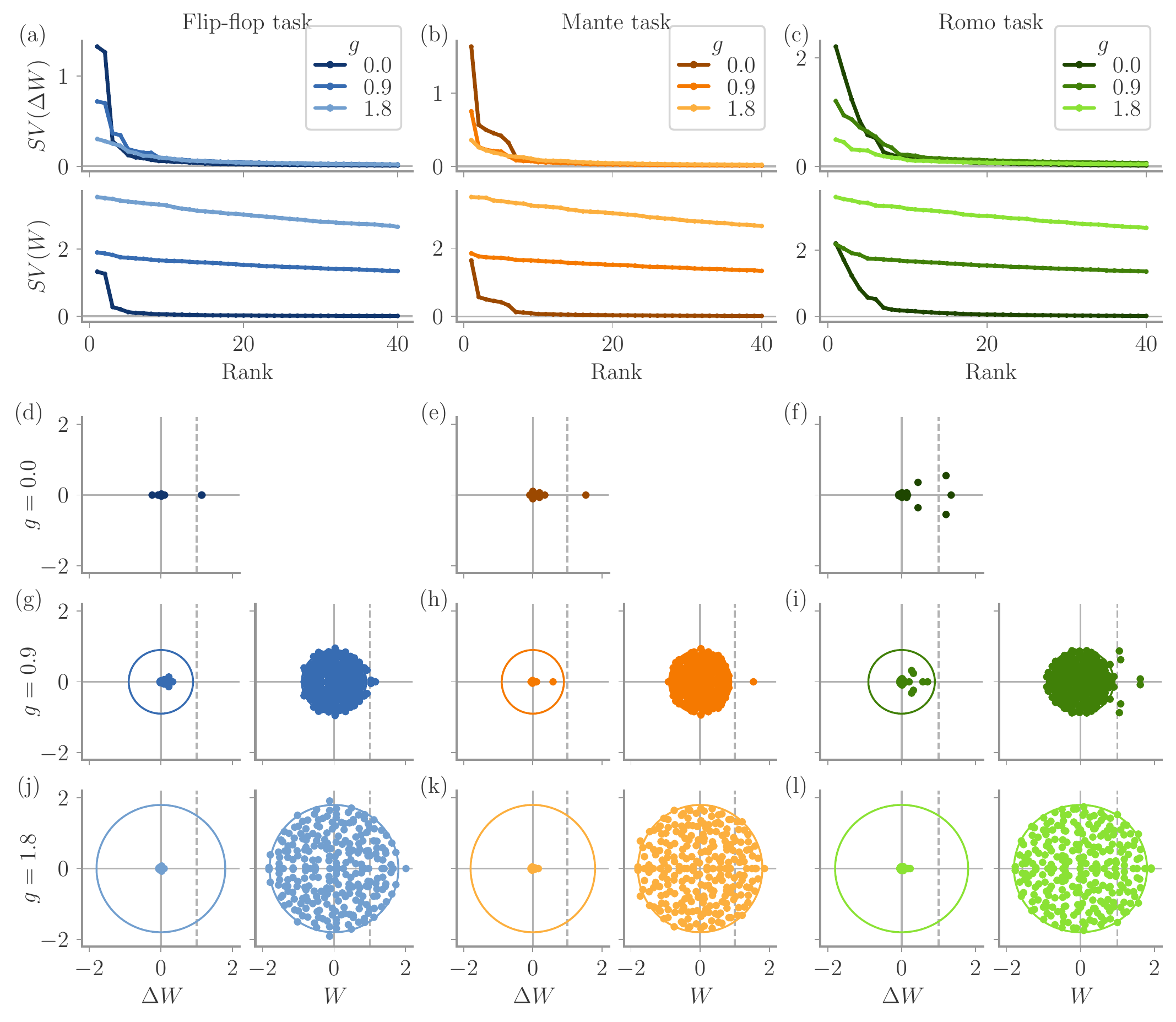}
  \caption{
  Singular values (SVs) and eigenvalues (EVs) of RNNs trained for all three tasks with different initial connectivity strength $g \in \{0.0, 0.9, 1.8\}$.
  \textbf{(a-c)} First 40 SVs of the weight changes $\Delta W$ (top) and the final weight matrix $W = W_0 + \Delta W$ (bottom). Note the different y-scales: 
  For $g=0$ (darkest lines), the SVs in both plots are the same. For larger $g$, the SVs of $\Delta W$ tend to become smaller, while those of $W$ increase.
  \textbf{(d-l)} Eigenvalue spectra for $\Delta W$ (left) and $W$ (right). The x- and y-coordinates are the real and imaginary part, respectively. 
  For $g=0$, \textbf{(d-f)}, the EVs of $\Delta W$ and $W$ are the same. 
  For $g>0$, we plot the circles with radius $g$ for comparison. Inside this radius, the eigenvalues of $W_0$ are distributed uniformly \protect\citeSM{ginibre1965statistical}. Note that most EVs of $W$ still remain with in this circle.
  Parameters as in Fig. 1 of the main text, specifically $N = 256$. 
  }
  \label{fig:all_tasks_sv_ev}
\end{figure}

\begin{figure}[htb]
  \centering
  \includegraphics[width=\figwidth]{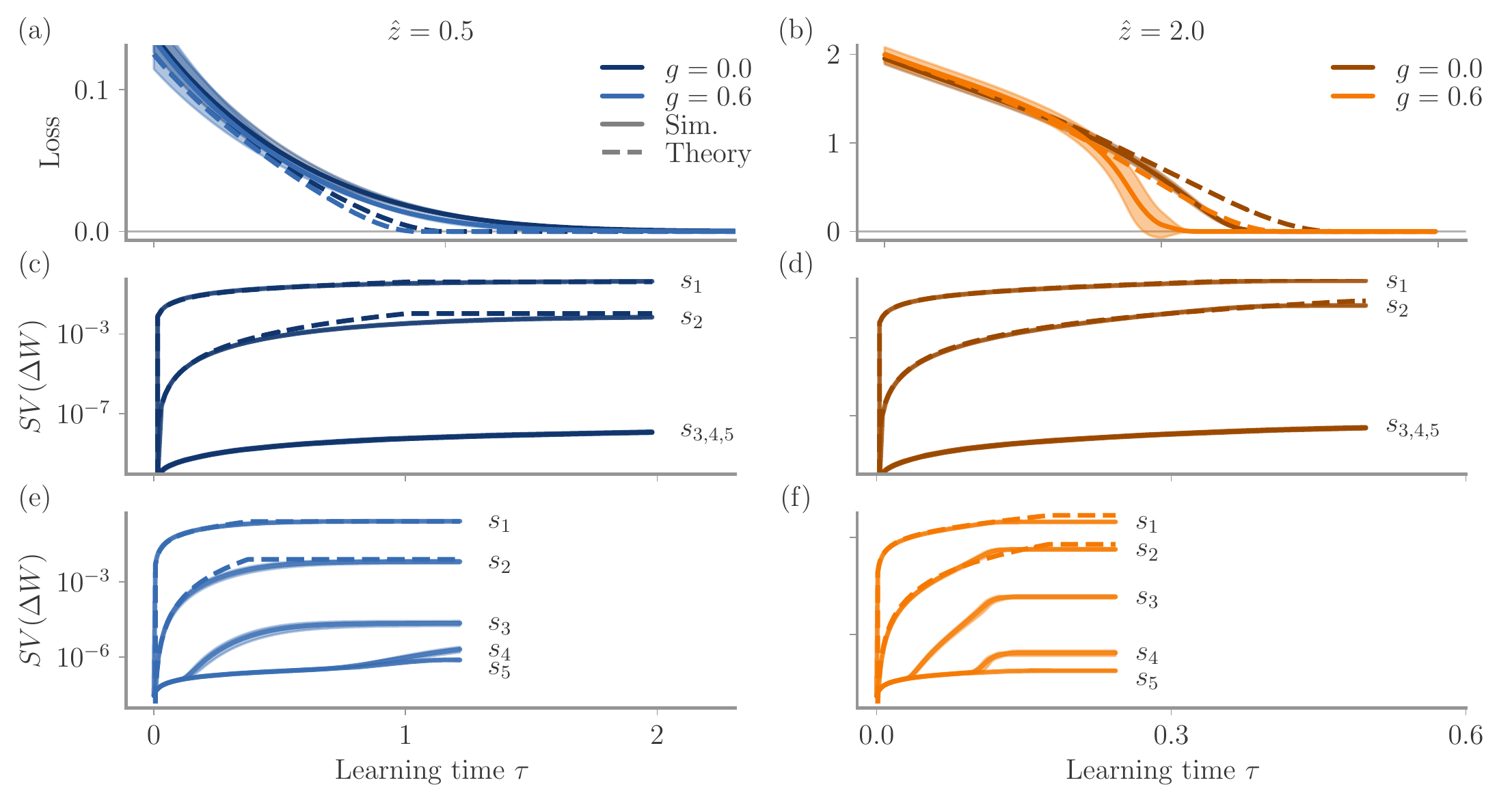}
  \caption{
  Evolution of SVs on log scale for the simple task, as a supplement to Fig. 3 of the main text. There, the SVs are shown on a linear scale, which does not allow to observe the evolution of any but the largest SVs. Our theory predicts only the first two SVs (dashed lines); any higher SVs are zero at order $\mathcal{O}(\tau^3)$.
  \textbf{(a,b)} Loss curves as a reference for the learning process. 
  \textbf{(c,d)} First five SVs for $g = 0$. Note that the curves of $s_3$, $s_4$, and $s_5$ overlap. 
  \textbf{(e,f)} First five SVs for $g=0.6$.
  }
  \label{fig:linear_loss_sv_log}
\end{figure}

\begin{figure}[htb]
  \centering
  \includegraphics[width=\textwidth]{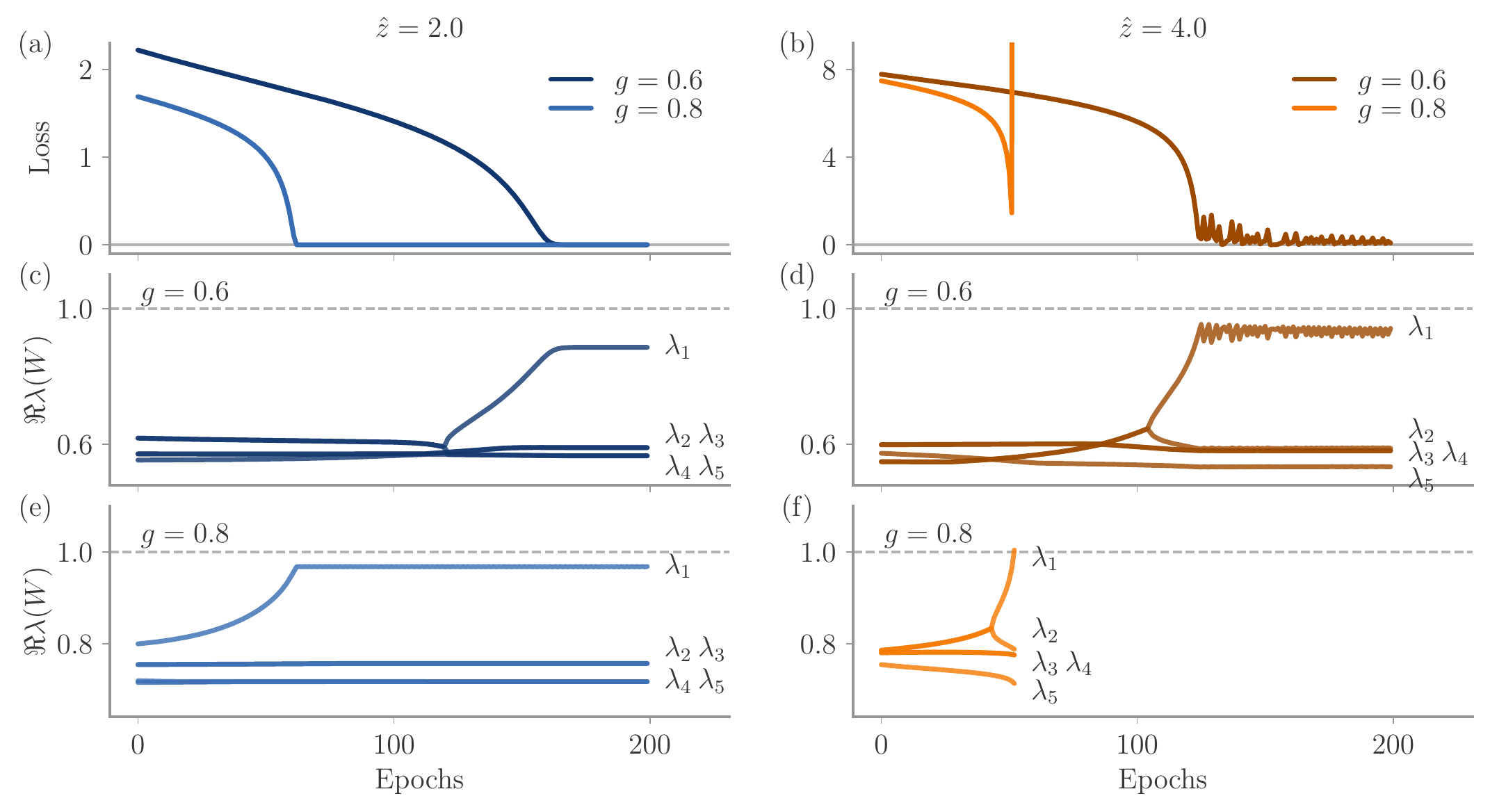}
  \caption{
  Exploding gradient when the real part of the largest eigenvalue $\lambda_1$ of $W$ crosses 1. For infinitely small learning rate $\eta$, the readout $z$ crosses the target value $\hat{z}$ before $\lambda_1$ crosses 1, so that learning stops. However, for a finite learning time, $z$ may become larger than an $\hat{z}$, and the divergent gradient may induce oscillations and failure of learning. This failure happens for large target values $\hat{z}$ and initial connectivity strength $g$, which promote the growth of $\lambda_1$.  
  \textbf{(a,b)} Loss curves for two different target values and initial connectivity strengths. For $\hat{z} = 4$ and $g=0.8$, the gradient diverges and learning stops.
  \textbf{(c-f)} Real parts of first five EVs $\lambda_i$ (order by decreasing real parts). Symbols at the end of each trajectory indicate the eigenvalues. In case of complex conjugates, the two corresponding $\lambda_i$ are written next to each other. The dashed grey line indicates the critical value $\Re \lambda = 1$ for which the gradients diverge. 
  Parameters: $N = 256$, $\eta = \eta_0 (1 - g^2)^2$ with $\eta_0 = 0.002,\, 0.001$ for $\hat{z} = 2,\, 4$, respectively.  
  Task parameters as in the main text but with longer trial time, $T = 201$ (so that the network still converges to the fixed point despite the slower time scales).
  }
  \label{fig:linear_failed_learning}
\end{figure}

\begin{figure}[htb]
  \centering
  \includegraphics[width=\textwidth]{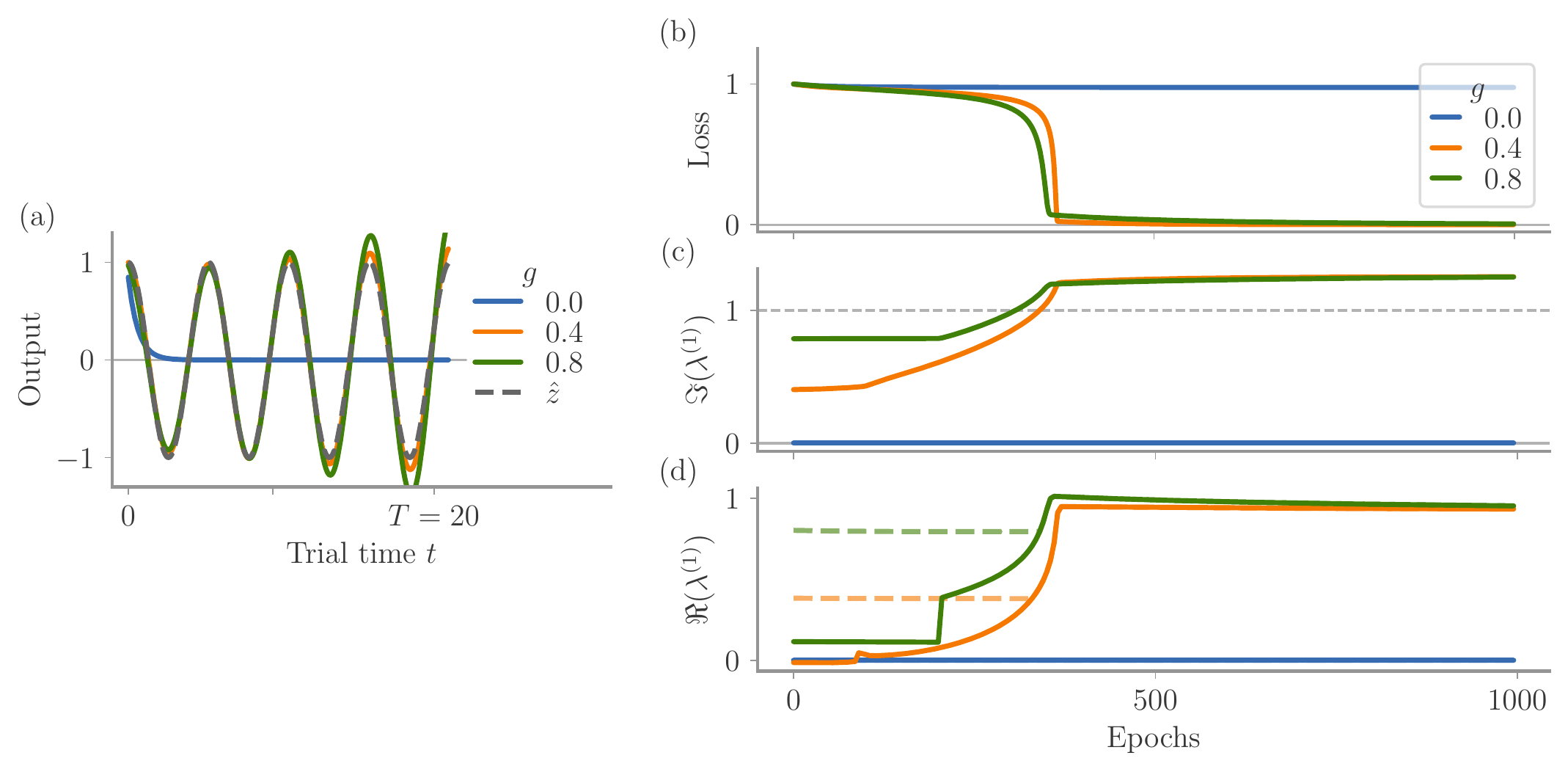}
  \caption{
  Example of learning only in presence of initial connectivity. For linear RNNs without initial connectivity, gradient descent-induced connectivity changes are always constructed from the input- and output vectors. If the space of these vectors is too small, learning fails. Here, we take the simple example of a linear network learning a cosine oscillation, starting from a fixed initial condition [see \textbf{(a)}]. The initial condition is set by a delta pulse through the input vector; otherwise, the input is zero. We set both input and output vector to $\mathbf{w}$, so that there is only a single vector available. However, creating the necessary complex conjugate eigenvalues needs a rank-two connectivity and hence at least two different directions. Random initial connectivity enlarges the pool of available directions. 
  \textbf{(a)} Output of networks at the end of training for three different values of $g$. Dashed line shows target 
  $\hat{z}(t) = \cos(2 \pi f t)$ with frequency $f=0.2$. Learning failed for $g=0$. For the other two values, the network finds a slightly unstable solution (perfect marginal stability is not achieved because of the limited trial time $T=20$). 
  \textbf{(b)} Loss over training epochs.
  \textbf{(c)} Imaginary part of largest eigenvalue $\lambda^{(1)}$, sorted by \textit{imaginary parts}. 
  \textbf{(d)} Real part of $\lambda^{(1)}$. The dashed lines show the real part of the largest eigenvalue sorted by real parts. 
  For $g=0$, no nonzero eigenvalue emerges throughout training. 
  Parameters: $N = 256$, $\eta = (0.2, 0.15, 0.05)$ for $g = (0.0, 0.4, 0.8)$, respectively (adapted heuristically for smooth convergence); training for 1000 epochs (batch size = 1, since there is not stochastic part). Simulation step size was reduced to $\Delta t = 0.1$. 
  }
  \label{fig:cos_loss_evs}
\end{figure}

\FloatBarrier

\section{Expansion of linear learning}
For the simple learning problem, the readout in the limit $t \to \infty$ is given by
\begin{equation} 
    \label{eq:fp_supp}
    z 
    = \mathbf{w}^T\! \left( I - W \right)^{-1} \mathbf{m}  \,.
\end{equation}
The loss is quadratic: $L = (\hat{z} - z)^2 / 2$. The weights change according to the gradient of the loss w.r.t. to recurrent weights $W$, namely
\begin{equation}
    \label{eq:grad_desc_full_supp}
    \frac{\mathrm{d} W(\tau) }{\mathrm{d} \tau}
    = -\frac{\mathrm{d}L}{\mathrm{d}W}
    = \left[\hat{z} - z(\tau) \right] 
    \, \left[ I - W^T\!(\tau) \right]^{-1}
    \mathbf{w} 
    \mathbf{m}^T
    \left[ I - W^T\!(\tau) \right]^{-1}  \,.
\end{equation}
We expand these dynamics in orders of $\tau$. In the main text, we introduced the expansion
\begin{equation}
    W(\tau) = \sum_{k=0}^{\infty} W_k \,\frac{\tau^k}{k!}  \,,
\end{equation}
with coefficients $W_k$ obtained from $\mathrm{d}^k W/\mathrm{d}\tau^k$ at $\tau = 0$. 

\subsection{First order}%
\label{sub:first_order}
Because of the independence of $W_0$, $\mathbf{w}$, and $\mathbf{m}$, the initial readout $z_0$ is zero, and we directly obtain
\begin{equation}
    W_1 = \hat{z} B^T\! \mathbf{w} \mathbf{m}^T\! B^T \,,
\end{equation}
with $B = (I - W_0)^{-1}$. The weight changes linear in $\tau$ are 
\begin{equation}
    \Delta W(\tau) = \mathbf{u}_1 \mathbf{v}_1^T + \mathcal{O}(\tau^2) \,,
\end{equation}
with 
\begin{equation}
    \mathbf{u}_1 = a_1 B^T\! \mathbf{w}  \,,\qquad
    \mathbf{v}_1^T = a_1 \mathbf{m}^T\! B^T  \,,
\end{equation}
and the coefficient 
\begin{equation}
    a_1^2 = \tau \hat{z} \,.
\end{equation}
Note that we chose to split the norm of the rank-one matrix equally between the two vectors, which simplifies notation later on. 
To compute the readout, we note that $W_1$ is a rank-one matrix. This allows us to apply the matrix inversion lemma (a.k.a. Sherman-Morrison formula; \cite{harville1998matrix}): 
The matrix $I - W_0$ is invertible for $g < 1$, and subtracting a rank-one matrix $\mathbf{u}\mathbf{v}^T$ changes its inverse to
\begin{equation}
    \label{eq:mil}
    \left( I - W_0 - \mathbf{u}\mathbf{v}^T \right)^{-1}
    =
    B + \frac{1}{1 - \mathbf{v}^T\! B \mathbf{u}} B \mathbf{u} \mathbf{v}^T\! B \,,
\end{equation}
To compute the readout at linear order, we first realize that the scalar product in the denominator in \cref{eq:mil} vanishes:
\begin{equation}
    \mathbf{v}_1^T\! B \mathbf{u}_1 
    = a_1^2 \mathbf{m}^T\! B^T\! B B^T \mathbf{w}   
    = 0 \,.
\end{equation}
To show this, we note that $\mathbf{m}$ and $\mathbf{w}$ are independent of $M = B^T\! B  B^T$, and therefore
\begin{equation}
    \mathbb{E} \left[\mathbf{m}^T\! M \mathbf{w} \right]
    = \sum_{i=1}^N \sum_{j=1}^N 
    \underbrace{\mathbb{E}[m_i w_j] }_{=0}
    \mathbb{E}[M_{ij}] \,.
\end{equation}
The variance of $\mathbf{m}^T\! M \mathbf{w}$ is of order $1/N$, so that in the limit of $N \to \infty$, the term self-averages to zero. 
With this, we can compute the readout:
\begin{equation}
\begin{split}
    z &= 
    \mathbf{w}^T\! 
    \left( I - W_0 - \mathbf{u}_1\mathbf{v}_1^T \right)^{-1} 
    \mathbf{m}
    \\ &=
    \underbrace{\mathbf{w}^T\! B \mathbf{m}}_{=0}
    + 
    \mathbf{w}^T\! 
    B \mathbf{u}_1 \, \mathbf{v}_1^T\! B 
    \mathbf{m}
    \\ &=
    \tau \hat{z} \,
    \mathbf{w}^T\! B B^T\! \mathbf{w} \,
    \mathbf{m}^T\! B^T\! B \mathbf{m}
    \\ &=
    \tau \hat{z} \beta^2 + \mathcal{O}(\tau^2) \,.
\end{split}
\end{equation}
The term $\mathbf{w}^T\!B B^T\!\mathbf{w}$ (and likewise 
$\mathbf{m}^T\!B^T\! B\mathbf{m}$) has expectation
\begin{equation}
    \mathbb{E} \left[\mathbf{w}^T\! B B^T\! \mathbf{w} \right]
    = \sum_{i=1}^N \sum_{j=1}^N 
    \underbrace{\mathbb{E}[w_i w_j] }_{=\delta_{ij}/N}
    \mathbb{E}[(B B^T)_{ij}] 
    = 
    \frac{1}{N} \mathbb{E}[\mathrm{Tr}(B B^T)] 
    = \beta
    \,.
\end{equation}
The expected trace $\beta = 1/ (1 - g^2)$ is computed in \cref{sec:traces}. 
Due to self-averaging in the limit $N \to \infty$, we omit the expectation. 

The singular values of $W_1$ are the square roots of the eigenvalues of 
\begin{equation}
    W_1 W_1^T
    = \hat{z}^2 
    B^T\! \mathbf{w} \mathbf{m}^T\! B^T
    B \mathbf{m} \mathbf{w}^T\! B \,.
\end{equation}
Since this is again a rank-one matrix, we compute the only nonzero eigenvalue via the trace:
\begin{equation}
    s^2
    = \mathrm{Tr}(W_1 W_1^T)
    = \hat{z}^2 
    \mathbf{w}^T\! B
    B^T\! \mathbf{w} \mathbf{m}^T\! B^T
    B \mathbf{m} 
    = \hat{z}^2 \beta^2 \,.
\end{equation}
The singular value, which is also the norm of $W_1$, is therefore
\begin{equation}
    s
    = ||W_1||
    = \hat{z} \beta \,.
\end{equation}
The learning time $\tau^*_1$ is the solution to the equation $z(\tau^*_1) = \hat{z}$, 
namely $\tau^*_1 = 1 / \beta^2$. The connectivity changes at this learning time are $\Delta W = \tau^*_1 W_1$, with norm $||\Delta W|| = \tau^*_1 ||W_1|| = \hat{z} / \beta$.

\subsection{Second order}%
\label{sub:second_order}
We again make use of the matrix inversion lemma, \cref{eq:mil}, and compute
\begin{equation}
    \label{eq:dW2dt}
    \begin{split}
    W_2
    &= 
    \frac{\mathrm{d}^2W}{\mathrm{d}\tau^2} 
    \bigg\rvert_{\tau = 0}
    \\&=
    \frac{\mathrm{d}}{\mathrm{d}\tau} 
    \left[
    (\hat{z}  - z) 
    \left( I - W_0 - \mathbf{u}_1\mathbf{v}_1^T \right)^{-T}
    \mathbf{w}\mathbf{m}^T\!
    \left( I - W_0 - \mathbf{u}_1\mathbf{v}_1^T \right)^{-T}
    \right]
    \bigg\rvert_{\tau = 0}
    \\&=
    \frac{\mathrm{d}}{\mathrm{d}\tau} 
    \left[
    (\hat{z}  - z) 
    B^T\!
    \left( I + \mathbf{v}_1 \mathbf{u}_1^T\! B^T\right)
    \mathbf{w} \mathbf{m}^T
    \left( I + B^T\! \mathbf{v}_1 \mathbf{u}_1^T \right)
    B^T
    \right]
    \bigg\rvert_{\tau = 0}
    \\&=
    \frac{\mathrm{d}}{\mathrm{d}\tau} 
    \left[
    (\hat{z}  - \tau \hat{z} \beta^2) 
    B^T\!
    \left( \mathbf{w} + \tau \hat{z} \beta B \mathbf{m} \right)
    \left( \mathbf{m}^T\! + \tau \hat{z} \beta \mathbf{w}^T\! B \right) 
    B^T\!
    \right]
    \bigg\rvert_{\tau = 0}
    \\&=
    \hat{z} \beta
    \, B^T\!
    \left[ 
        - \beta \mathbf{w} \mathbf{m}^T
        +
        \hat{z} 
        \left( \mathbf{w} \mathbf{w}^T\! B + B \mathbf{m} \mathbf{m}^T \right) 
    \right]
    B^T\! \,.
    \end{split}
\end{equation}
We notice that the weight changes up to order $\mathcal{O}(\tau^2)$ can be written as the outer product of two vectors and is thus a rank-one matrix: 
\begin{equation}
    \begin{split}
    \Delta W &= \tau W_1 + \frac{\tau^2}{2} W_2 + \mathcal{O}(\tau^3)
    \\&=
    B^T\!
    \left[ 
    \left(\tau \hat{z} - \frac{\tau^2}{2} \hat{z} \beta^2 \right)
        \mathbf{w} \mathbf{m}^T
        +
        \frac{\tau^2}{2} \hat{z}^2 \beta 
        \left( \mathbf{w} \mathbf{w}^T\! B + B \mathbf{m} \mathbf{m}^T \right) 
    \right]
    B^T + \mathcal{O}(\tau^3) 
    \\&=
    B^T\!
    \left( a_2 \mathbf{w} + b_2 B \mathbf{m} \right)
    \left( a_2 \mathbf{m}^T + b_2 \mathbf{w}^T\!B^T\! \right)
    B^T 
    + \mathcal{O}(\tau^3)  
    \\&= \mathbf{u}_2 \mathbf{v}_2^T
    + \mathcal{O}(\tau^3)   \,,
    \end{split}
\end{equation}
with 
\begin{equation}
    \mathbf{u}_2 
    = 
    B^T\!
    \left( a_2 \mathbf{w} + b_2 B \mathbf{m} \right)
    \,,\qquad
    \mathbf{v}_2^T = 
    \left( a_2 \mathbf{m}^T + b_2 \mathbf{w}^T\!B^T\! \right)
    B^T  \,.
\end{equation}
The coefficients are implicitly defined by
\begin{equation}
    a_2^2 = \tau \hat{z} - \frac{\tau^2}{2}\hat{z} \beta^2 \,,\qquad
    a_2 b_2 = \frac{\tau^2}{2} \hat{z}^2 \beta \,.
\end{equation}
Note that the correction $b_2^2$ from completing the square is of order $\mathcal{O}(\tau^3)$.

Similarly to the first order, we can compute the readout $z$:
\begin{equation}
    z_2 
    = 
    \frac{
    \mathbf{w}^T\! B \mathbf{u}_2
    \mathbf{v}_2^T\! B \mathbf{m}
    }{1 - \mathbf{v}_2^T\! B \mathbf{u}_2} 
    = 
    a_2^2 \beta^2 + \mathcal{O}(\tau^3) \,,
\end{equation}
with 
\begin{equation}
    \mathbf{w}^T\! B \mathbf{u}_2 
    = \mathbf{v}_2^T\! B \mathbf{m} 
    = a_2 \beta  \,.
\end{equation}
The denominator is of order $\mathcal{O}(\tau^2)$ and hence 
does not contribute to $z_2$:
\begin{equation}
    \begin{split}
    \mathbf{v}_2^T\! B \mathbf{u}_2 
    &= 
    \left( a_2 \mathbf{m}^T + b_2 \mathbf{w}^T\!B^T\! \right)
    B^T\! B B^T\!
    \left( a_2 \mathbf{w} + b_2 B \mathbf{m} \right)
    \\&=
    2 a_2 b_2 \gamma + \mathcal{O}(\tau^3)  \,.
    \end{split}
\end{equation}
The random matrix term 
$\gamma = \mathbf{w}^T\! B B^T\! B B^T\! \mathbf{w} =\beta^4$ is 
compute \cref{sec:traces}. Terms of the form $\mathbf{m}^T\! M \mathbf{w}$, with $M$ constructed from $B$ and $B^T$ are zero due to the independence of all three quantities. 

\subsection{Third order}%
\label{sub:third_order}
Since $\Delta W$ at order $\mathrm{O}(\tau^2)$ is a rank-1 matrix, we can use the same formalism as for the second order, cf. \cref{eq:dW2dt}. We now only keep terms with $\tau^2$:
\begin{equation}
    \label{eq:dW3dt}
    \begin{split}
    W_3
    &= 
    \frac{\mathrm{d}^3W}{\mathrm{d}\tau^3} 
    \bigg\rvert_{\tau = 0}
    \\&=
    \frac{\mathrm{d}^2}{\mathrm{d}\tau^2} 
    \left[
    (\hat{z}  - z) 
    \left( I - W_0 - \mathbf{u}_2\mathbf{v}_2^T \right)^{-T}
    \mathbf{w}\mathbf{m}^T\!
    \left( I - W_0 - \mathbf{u}_2\mathbf{v}_2^T \right)^{-T}
    \right]
    \bigg\rvert_{\tau = 0}
    \\&=
    \frac{\mathrm{d}^2}{\mathrm{d}\tau^2} 
    \left[
    (\hat{z}  - z) 
    B^T\!
    \left( I + \mathbf{v}_2 \mathbf{u}_2^T\! B^T\right)
    \mathbf{w} \mathbf{m}^T
    \left( I + B^T\! \mathbf{v}_2 \mathbf{u}_2^T \right)
    B^T\!
    \right]
    \bigg\rvert_{\tau = 0}
    \\&=
    \frac{\mathrm{d}^2}{\mathrm{d}\tau^2} 
    \left[
    (\hat{z}  - a_2^2 \beta^2) 
    B^T\!
    \left[ \mathbf{w} + \beta B \left( a_2^2 \mathbf{m} + a_2 b_2 B^T\! \mathbf{w}\right)\right]
    \left[ \mathbf{m}^T\! + \beta \left( a_2^2 \mathbf{w}^T\! + a_2 b_2 \mathbf{m}^T\! B^T \right) B \right]
    B^T\!
    \right]
    \bigg\rvert_{\tau = 0}
    \\&=
    \hat{z}
    \beta^2
    B^T\! \left[ 
    \beta^2
    \mathbf{w} \mathbf{m}^T\!
    - 3
    \hat{z} \beta
    \left( \mathbf{w} \mathbf{w}^T\! B + B \mathbf{m} \mathbf{m}^T\!\right)
    + 
    2 \hat{z}^2 
    B \mathbf{m} \mathbf{w}^T\! B 
    + 
    \hat{z}^2 
    \left( \mathbf{w} \mathbf{m}^T\! B^T\! B + B B^T\! \mathbf{w} \mathbf{m}^T\!\right)
    \right] B^T\! \,.
    \end{split}
\end{equation}
The changes up to order $\mathcal{O}(\tau^2)$ are now of rank two:
\begin{equation}
    \label{eq:dW3}
    \begin{split}
    \Delta W &= \tau W_1
    + \frac{\tau^2}{2} W_2 
    + \frac{\tau^3}{6} W_2 
    + \mathcal{O}(\tau^4)
    \\&=
    B^T\!
    \Bigg[ 
    \left(\tau \hat{z} - \frac{\tau^2}{2} \hat{z} \beta^2  + \frac{\tau^3}{6} \hat{z} \beta^4 \right) 
    \mathbf{w} \mathbf{m}^T
    + \left(\frac{\tau^2}{2} \hat{z}^2 \beta - \frac{\tau^3}{2} \hat{z}^2 \beta^3 \right) 
    \left( \mathbf{w} \mathbf{w}^T\! B + B \mathbf{m} \mathbf{m}^T \right) 
    \\&\qquad 
    +
    \frac{\tau^3}{3} \hat{z}^3 \beta^2 
    B \mathbf{m} \mathbf{w}^T\! B 
    +
    \frac{\tau^3}{6} \hat{z}^3 \beta^2 
    \left( \mathbf{w} \mathbf{m}^T\! B^T\! B + B B^T\! \mathbf{w} \mathbf{m}^T\!\right)
    \Bigg]
    B^T 
    + \mathcal{O}(\tau^4) 
    \\&=
    B^T\!
    \left( a_3 \mathbf{w} + b_3 B \mathbf{m} + c_3 B B^T\! \mathbf{w} \right)
    \left( a_3 \mathbf{m}^T + b_3 \mathbf{w}^T\!B^T\! + c_3 \mathbf{m}^T\! B^T\! B \right)
    B^T + 
    \hat{b}_3^2 
    B^T\! B \mathbf{m} \mathbf{w}^T\! B B^T\!
    + \mathcal{O}(\tau^4)  
    \\&= \mathbf{u}_3 \mathbf{v}_3^T
    + \mathbf{\hat{u}}_3 \mathbf{\hat{v}}_3^T 
    + \mathcal{O}(\tau^4)  
    \,,
    \end{split}
\end{equation}
with
\begin{align}
    \mathbf{u}_3 
    &=
    B^T\! \left(a_3 \mathbf{w} + b_3 B \mathbf{m} + c_3 B B^T\! \mathbf{w} \right)
    \,,\\
    \mathbf{v}_3^T 
    &=
    \left(a_3 \mathbf{m}^T\! + b_3 \mathbf{w}^T\! B + c_3 \mathbf{m}^T\! B^T\! B \right) B^T\!
    \,,\\
    \mathbf{\hat{u}}_3 
    &=
    \hat{b}_3
    B^T\! B \mathbf{m} 
    \,,\\
    \mathbf{\hat{v}}_3^T 
    &=
    \hat{b}_3
    \mathbf{w}^T\! B B^T \,.
\end{align}
The coefficients are implicitly defined by 
\begin{align}
    \label{eq:a3}
    a_3^2 &= \tau \hat{z} - \frac{\tau^2}{2}\hat{z} \beta^2 + \frac{\tau^3}{6} \hat{z} \beta^4 \,,\\
    \label{eq:a3b3}
    a_3 b_3 &= \frac{\tau^2}{2} \hat{z}^2 \beta - \frac{\tau^3}{2} \hat{z}^2 \beta^3 \,,\\
    \label{eq:a3c3}
    a_3 c_3 &= \frac{\tau^3}{6} \hat{z}^3 \beta^2 
    \,,\\
    \label{eq:b3}
    b_3^2 &= 
    \frac{(a_3 b_3)^2}{a_3^2}
    = \frac{\tau^3}{4} \hat{z}^3 \beta^2
    \,,\\
    \label{eq:hatb3}
    \hat{b}_3^2 
    &= \frac{\tau^3}{3} \hat{z}^3 \beta^2  - b_3^2 
    = \frac{\tau^3}{12} \hat{z}^3 \beta^2 \,.
\end{align}
The remaining corrections $b_3 c_3$ and $c_3^2$ are of order $\mathcal{O}(\tau^4)$ or higher. 

The changes $\Delta W$ can be written in a compact rank-two form:
\begin{equation}
    \Delta W(\tau)
    =
    \begin{bmatrix}
    \mathbf{u}_3 & \mathbf{\hat{u}}_3
    \end{bmatrix}
    \begin{bmatrix}
    \mathbf{v}_3^T \\ \mathbf{\hat{v}}_3^T 
    \end{bmatrix}  
    + \mathcal{O}(\tau^4) 
    = U V^T 
    + \mathcal{O}(\tau^4) 
    \,.
\end{equation}
With this, we compute the readout, using the matrix inversion lemma \cite{harville1998matrix}:
\begin{equation}
\label{eq:z3}
\begin{split}
    z 
    &= \mathbf{w}^T\! \left( I - W_0 - U V^T \right)^{-1} \mathbf{m} 
    + \mathcal{O}(\tau^4) 
    \\&= \mathbf{w}^T\!
    \left[ B + B U \left( I_2 - V^T\! B U \right)^{-1} V^T\! B \right] \mathbf{m} 
    + \mathcal{O}(\tau^4) 
    \\&= \mathbf{w}^T\! B U \left( I_2 - V^T\! B U \right)^{-1} V^T\! B \mathbf{m} 
    + \mathcal{O}(\tau^4) 
    \,.
\end{split}
\end{equation}
Here, $I_2$ is the $2\times2$ identity matrix. We compute the entries of $V^T\! B U$ up to $\mathcal{O}(\tau^3)$:
\begin{align}
    \mathbf{v}_3^T\! B \mathbf{u}_3 
    &= 2 a_3 b_3 \gamma  \,,\\ 
    \mathbf{v}_3^T\! B \mathbf{\hat{u}}_3 
    &= a_3 \hat{b}_3 \gamma  \,,\\ 
    \mathbf{\hat{v}}_3^T\! B \mathbf{u}_3 
    &= a_3 \hat{b}_3 \gamma  \,,\\
    \mathbf{\hat{v}}_3^T\! B \mathbf{\hat{u}}_3 
    &= 0 \,.
\end{align}
The factor $\gamma=\beta^4$ is computed in \cref{sec:traces}.
Therefore, 
\begin{equation}
    I_2 - V^T\! B U
    =
    \begin{bmatrix}
    1 - x & -y  \\
    -x & 1
    \end{bmatrix} \,,
\end{equation}
with 
$
x = \mathbf{v}_3^T\! B \mathbf{u}_3  \,,
$
and
$
y = \mathbf{v}_3^T\! B \mathbf{\hat{u}}_3  
$.
Since $p$ and $q$ are $\mathcal{O}(\tau^2)$, we have
\begin{equation}
    \left( I_2 - V^T\! B U \right)^{-1} 
    = \frac{1}{1 - x - y^2} 
    \begin{bmatrix}
    1 & y  \\
    y & 1 - x
    \end{bmatrix}
    =
    \begin{bmatrix}
    1 + x & y  \\
    y & 1 
    \end{bmatrix}
    + \mathcal{O}(\tau^4) \,.
\end{equation}
To complete the evaluation of $z$, \cref{eq:z3}, we further compute $\mathbf{w}^T\! B U$ and $V^T\! B \mathbf{m}$:
\begin{align}
    \mathbf{w}^T\! B \mathbf{u}_3 
    &= \mathbf{v}_3^T\! B \mathbf{m}
    = a_3 \beta + c_3 \gamma \,,\\
    \mathbf{w}^T\! B \mathbf{\hat{u}}_3 
    &= \mathbf{\hat{v}}_3^T\! B \mathbf{m}
    = 0 \,.
\end{align}
Hence, 
\begin{equation}
    \label{eq:z_final}
    \begin{split}
    z
    &= 
    \begin{bmatrix}
        \mathbf{w}^T\! B \mathbf{u}_3  
        &
    \mathbf{w}^T\! B \mathbf{\hat{u}}_3 
    \end{bmatrix}
    \begin{bmatrix}
    1 + x & y  \\
    y & 1 
    \end{bmatrix}
    \begin{bmatrix}
        \mathbf{v}_3^T\! B \mathbf{m} \\
        \mathbf{\hat{v}}_3^T\! B \mathbf{m}
    \end{bmatrix}
    + \mathcal{O}(\tau^4) 
    \\&=
    (1 + x) 
    \mathbf{w}^T\! B \mathbf{u}_3 \, \mathbf{v}_3^T\! B \mathbf{m}
    + \mathcal{O}(\tau^4) 
    \\&=
    (1 + 2 a_3 b_3 \gamma) \,(a_3 \beta + c_3 \gamma)^2
    + \mathcal{O}(\tau^4) 
    \\&=
    \Big(1 + 
    \underbrace{2 a_3 b_3 \gamma}_{\mathcal{O}(\tau^2)} \Big) \, \Big(
    \underbrace{a_3^2 \beta^2}_{\mathcal{O}(\tau)}
    + \underbrace{2 a_3 c_3 \beta \gamma}_{\mathcal{O}(\tau^3)} 
    + \underbrace{c_3^2 \gamma^2}_{\mathcal{O}(\tau^4)} \Big)
    + \mathcal{O}(\tau^4) 
    \\&=
    a_3^2 \beta^2
    + 2 a_3 c_3 \beta \gamma
    + 2 a_3^2 a_3 b_3 \beta^2 \gamma
    + \mathcal{O}(\tau^4) 
    \\&= \hat{z} 
    \left[
    \beta^2 \tau 
    -  \frac{(\beta^2 \tau)^2}{2}
    + (1 + 8 \hat{z}^2 \beta) \frac{(\beta^2 \tau)^3}{6}
    \right] 
    + \mathcal{O}(\tau^4) 
    \,.
    \end{split}
\end{equation}
The last lines are based on the implicit definitions of the coefficients $a_3$, $b_3$, and $c_3$ in \cref{eq:a3,eq:a3b3,eq:a3c3} and $\gamma = \beta^4$.

We end this section with looking at the special case $g = 0$. 
With $B = I$ and $\beta = 1$, the weight changes \cref{eq:dW3} simplify to
\begin{equation}
    \begin{split}
    \Delta W 
    &=
    \left(\tau \hat{z} - \frac{\tau^2}{2} \hat{z}   + \frac{\tau^3}{2} \hat{z} \right) 
    \mathbf{w} \mathbf{m}^T
    + \left(\frac{\tau^2}{2} \hat{z}^2  - \frac{\tau^3}{2} \hat{z}^2  \right) 
    \left( \mathbf{w} \mathbf{w}^T\! + \mathbf{m} \mathbf{m}^T \right) 
    +
    \frac{\tau^3}{3} \hat{z}^3  
    \mathbf{m} \mathbf{w}^T\!
    + \mathcal{O}(\tau^4) 
    \\&=
    \begin{bmatrix}
    \mathbf{w} & \mathbf{m}
    \end{bmatrix}
    \begin{bmatrix}
    A_{11} & A_{12} \\ A_{21} & A_{22}
    \end{bmatrix}
    \begin{bmatrix}
    \mathbf{w}^T \\ \mathbf{m}^T
    \end{bmatrix}
    \,,
    \end{split}
\end{equation}
with
\begin{align}
    A_{11} &= \frac{\hat{z}^2}{2} \left(\tau^2  - \tau^3\right) 
    + \mathcal{O}(\tau^4) 
    \,, \\
    A_{12} &= \hat{z} \left(\tau - \frac{\tau^2}{2} + \frac{\tau^3}{6} (1 + 2 \hat{z}^2) \right) 
    + \mathcal{O}(\tau^4) 
    \,, \\
    A_{21} &= \frac{\hat{z}^3\tau^3}{3} 
    + \mathcal{O}(\tau^4) 
    \,,
\end{align}
and $A_{22} = A_{11}$.
Note that for $g=0$, one can write the entire gradient descent dynamics in terms of the matrix $2\times 2$ matrix $A$:
\begin{equation}
    \frac{\mathrm{d}A}{\mathrm{d}\tau}
    = 
    (\hat{z} - z) 
    \left[I + C^T\right]
    \begin{bmatrix}
    1 \\ 0
    \end{bmatrix}
    \begin{bmatrix}
    0 & 1
    \end{bmatrix}
    \left[I + C^T\right] \,,
\end{equation}
with 
\begin{equation}
    z = 
    \begin{bmatrix}
    1 & 0
    \end{bmatrix}
    \left[I + C\right]
    \begin{bmatrix}
    0 \\ 1
    \end{bmatrix} 
    =
    C_{12}
    \,,
\end{equation}
and
\begin{equation}
    C = A (I - A)^{-1} \,.
\end{equation}
With the symmetry $A_{11} = A_{22}$, this equation still has three degrees of freedom, and we were not able to find a closed form solution.

\subsection{Singular values of weight changes}
The singular values of $\Delta W$ are determined by the eigenvalues of $\Delta W^T\! \Delta W$ up to order $\mathcal{O}(\tau^3)$. For the rank-two matrix $\Delta W = UV^T$, these are the eigenvalues of the matrix
\begin{equation}
    P
    = V^T\!V U^T\!U
    =
    \begin{bmatrix}
    p & q \\ q & r
    \end{bmatrix}^2
    =
    \begin{bmatrix}
    p^2 + q^2 & q(p + r) \\ q(p + r) & q^2 + r^2
    \end{bmatrix}^2 \,.
\end{equation}
As before, we compute the coefficients up to order $\mathcal{O}(\tau^3)$:
\begin{align}
    p &= \mathbf{u}_3^T\!\mathbf{u}_3
    = \mathbf{v}_3^T\!\mathbf{v}_3
    = a^2 \beta + (b^2 + 2 ac) \gamma 
    \,,\\ 
    q &= \mathbf{u}_3^T\!\mathbf{\hat{u}}_3
    = \mathbf{v}_3^T\!\mathbf{\hat{v}}_3  
    =b \hat{b} \gamma
    \,,\\ 
    r &= \mathbf{\hat{u}}_3^T\!\mathbf{\hat{u}}_3
    = \mathbf{\hat{v}}_3^T\!\mathbf{\hat{v}}_3 
    = \hat{b}^2 \gamma \,.
\end{align}
The squared singular values are therefore
\begin{equation}
    s_\pm^2 = \frac{1}{2} \left(\mathrm{Tr}P \pm \sqrt{(\mathrm{Tr}P)^2 - 4 |P|}\right) \,.
\end{equation}
The terms are of order $p = \mathcal{O}(\tau)$ and $q, r = \mathcal{O}(\tau^3)$, so that 
\begin{align}
    \mathrm{Tr} &= p^2 + 2 q^2 + r^2 = \mathcal{O}(\tau^2) \,,\\
    |P| &= (pr - q^2)^2 = \mathcal{O}(\tau^8) \,.
\end{align}
This means that the solutions have different orders:
\begin{align}
    s_+^2 &= \mathrm{Tr}P - \frac{|P|}{\mathrm{Tr}P} \,,\\
    s_-^2 &= \frac{|P|}{\mathrm{Tr}P} \,.
\end{align}
Taking the square roots and sorting out the orders yields a linear first singular value, 
\begin{equation}
    s_+ 
    = \frac{\hat{z}}{\beta}
        \left[
        \beta^2 \tau
        - \frac{(\beta^2 \tau)^2}{2}  
        + \left(1 + \frac{7}{2} \hat{z}^2 \beta
        \right) \frac{(\beta^2 \tau)^3}{6}  \right]\,.
\end{equation}
The second singular value is cubic in learning time:
\begin{equation}
        s_-
        = \hat{b}_{(3)}^2 \gamma
        = \hat{z}^3\frac{(\beta^2 \tau)^3}{12} \,.
\end{equation}

\subsection{Effect of shuffling}%
Shuffling $W_0$ at the end of training destroys any correlation between $W_0$ and $W_1$, while keeping the same statistics. We denote that shuffled $W_0$ by $W_0^s$, and the corresponding inverse by $B^s = (1 - W_0^s)^{-1}$. 

At first order, the shuffled readout is 
\begin{equation}
\begin{split}
    \label{eq:z_s}
    z^s(\tau) &= 
    \mathbf{w}^T\! 
    \left( I - W_0^s - \tau^*_1 W_1 \right)^{-1} 
    \mathbf{m}
    \\ &=
    \mathbf{w}^T\! 
    \Big[
    B^s + \frac{1}{1 - \underbrace{\mathbf{v}_1^T\! B^s \mathbf{u}_1}_{=0}} B^s \mathbf{u}_1 \mathbf{v}_1^T\! B^s 
    \Big]
    \mathbf{m}
    \\ &=
    \underbrace{\mathbf{w}^T\! B^s \mathbf{m}}_{=0}
    + 
    \mathbf{w}^T\! B^s \mathbf{u}_1 \, \mathbf{v}_1^T\! B^s \mathbf{m}
    \\ &=
    \tau \hat{z} \,
    \mathbf{w}^T\! B^s B^T\! \mathbf{w} \,
    \mathbf{m}^T\! B^T\! B^s \mathbf{m}
    \\ &=
    \tau \hat{z} + \mathcal{O}(\tau^2) \,.
\end{split}
\end{equation}
The factor $\beta$ vanishes because
\begin{equation}
    \mathbb{E} \left[\mathbf{w}^T\! B^s B^T\! \mathbf{w} \right]
    = 
    \sum_{i=1}^N 
    \sum_{j=1}^N 
    \sum_{k=1}^N 
    \underbrace{\mathbb{E}[w_i w_k] }_{=\delta_{ik}/N}
    \mathbb{E}[B^s_{ij}] 
    \mathbb{E}[B^T_{jk}] 
    = 
    \frac{1}{N} 
    \sum_{i=1}^N 
    \sum_{j=1}^N 
    \underbrace{\mathbb{E}[B^s_{ij}]}_{=\delta_{ij} \left(1 + \frac{1}{N}\right)}
    \underbrace{\mathbb{E}[B^T_{ji}]}_{=\delta_{ji} \left(1 + \frac{1}{N}\right)}
    = 1 + \mathcal{O}(1/N)
    \,.
\end{equation}
Inserting $\tau^*_1 = 1/\beta^2$ into \cref{eq:z_s} yields $z^s(\tau^*_1) = \hat{z} / \beta^2$. The corresponding loss is 
\begin{equation}
    L^s 
    = \frac{1}{2} (\hat{z} - z^s(\tau^*_1))^2
    = \frac{1}{2} \hat{z}^2 \left(1 - \frac{1}{\beta^2}\right)^2
    = L_0 g^4 (2 - g^2)^2  \,,
\end{equation}
with initial loss $L_0 = \hat{z}^2/2$.

For the third order, not all amplification is lost:
Replacing $B$ with $B^s$ in the evaluation of $z$, 
\cref{eq:z3} yields
\begin{equation}
\label{eq:z3s}
\begin{split}
    z^s 
    &= \mathbf{w}^T\! \left( I - W_0^s - U V^T \right)^{-1} \mathbf{m} 
    + \mathcal{O}(\tau^4) 
    \\&= \mathbf{w}^T\! B^s U \left( I_2 - V^T\! B^s U \right)^{-1} V^T\! B^s \mathbf{m} 
    + \mathcal{O}(\tau^4) 
    \,.
\end{split}
\end{equation}
We compute 
\begin{equation}
    x^s 
    = \mathbf{v}_3^T\! B^s \mathbf{u}_3 
    = a_3 b_3 \left(
    \mathbf{m}^T\! B^T\! B^s B^T\! B \mathbf{m} +
    \mathbf{w}^T\! B B^T\! B^s B^T\! \mathbf{w}
    \right)
    = 2 a_3 b_3 \beta^2 \,.
\end{equation}
This is based on 
\begin{equation}
\begin{split}
    \mathbb{E} \left[\mathbf{w}^T\! B B^T\! B^s B^T\! \mathbf{w} \right]
    &= 
    \sum_{i=1}^N 
    \sum_{j=1}^N 
    \underbrace{\mathbb{E}[w_i w_j] }_{=\delta_{ij}/N}
    \mathbb{E}[(B B^T\! B^s B^T)_{ij}] 
    \\&= 
    \frac{1}{N}
    \sum_{i, j, k, l}
    \mathbb{E}[B_{ij} B^T_{jk} B^T_{li}] 
    \underbrace{\mathbb{E}[B^s_{kl}]}_{=\delta_{kl} \left(1 + \frac{1}{N}\right)}
    \\&= 
    \frac{1}{N}
    \sum_{i, j, k, l}
    \mathbb{E}[B_{ij} B^T_{jk} B^T_{ki}] 
    \underbrace{\mathbb{E}[B^s_{kl}]}_{=\delta_{kl} \left(1 + \frac{1}{N}\right)}
    \\&= 
    \frac{1}{N} 
    \mathbb{E}[\mathrm{Tr}( B B^T B^T)] 
    = \beta^2
    \,.
\end{split}
\end{equation}
Similarly, 
\begin{equation}
    y^s 
    = \mathbf{v}_3^T\! B^s \mathbf{\hat{u}}_3 
    = \mathbf{\hat{v}}_3^T\! B^s \mathbf{u}_3 
    = a_3 \hat{b}_3 \beta^2  \,,\qquad
    \mathbf{\hat{v}}_3^T\! B^s \mathbf{\hat{u}}_3 
    = 0 \,,
\end{equation}
and
\begin{align}
    \mathbf{w}^T\! B^s \mathbf{u}_3 
    &= \mathbf{v}_3^T\! B^s \mathbf{m}
    = a_3 + c_3 \beta^2 \,,\\
    \mathbf{w}^T\! B^s \mathbf{\hat{u}}_3 
    &= \mathbf{\hat{v}}_3^T\! B^s \mathbf{m}
    = 0 \,.
\end{align}
The remaining parts of the calculation of $z$ are similar to the case without shuffling, and the corresponding result to \cref{eq:z_final} is:
\begin{equation}
    \label{eq:z_s3}
    \begin{split}
    z^s
    &= 
    \begin{bmatrix}
        \mathbf{w}^T\! B^s \mathbf{u}_3  
        &
    \mathbf{w}^T\! B^s \mathbf{\hat{u}}_3 
    \end{bmatrix}
    \begin{bmatrix}
    1 + x^s & y^s  \\
    y^s & 1 
    \end{bmatrix}
    \begin{bmatrix}
        \mathbf{v}_3^T\! B^s \mathbf{m} \\
        \mathbf{\hat{v}}_3^T\! B^s \mathbf{m}
    \end{bmatrix}
    + \mathcal{O}(\tau^4) 
    \\&=
    (1 + x^s) 
    \mathbf{w}^T\! B^s \mathbf{u}_3 \, \mathbf{v}_3^T\! B^s \mathbf{m}
    + \mathcal{O}(\tau^4) 
    \\&=
    (1 + 2 a_3 b_3 \beta^2) \,(a_3 + c_3 \beta^2)^2
    + \mathcal{O}(\tau^4) 
    \\&=
    \Big(1 + 
    \underbrace{2 a_3 b_3 \beta^2}_{\mathcal{O}(\tau^2)} \Big) \, \Big(
    \underbrace{a_3^2}_{\mathcal{O}(\tau)}
    + \underbrace{2 a_3 c_3 \beta^2}_{\mathcal{O}(\tau^3)} 
    + \underbrace{c_3^2 \beta^4}_{\mathcal{O}(\tau^4)} \Big)
    + \mathcal{O}(\tau^4) 
    \\&=
    a_3^2
    + 2 a_3 c_3 \beta^2
    + 2 a_3^2 a_3 b_3 \beta^2 
    + \mathcal{O}(\tau^4) 
    \\&= \frac{\hat{z} }{\beta^2}
    \left[
    \beta^2 \tau 
    -  \frac{(\beta^2 \tau)^2}{2}
    + \left(1 + 2 \hat{z}^2 \left(1 + \frac{3}{\beta}\right)\right) \frac{(\beta^2 \tau)^3}{6}
    \right] 
    + \mathcal{O}(\tau^4) 
    \,.
    \end{split}
\end{equation}
A comparison with \cref{eq:z_final} shows that the first and second order terms are decreased by $1/\beta^2$. However, the third order term has a correction to this, similar to the learning time $\tau^*$.

\section{Traces}
\label{sec:traces}
Here we compute traces appearing in our learning problem:
\begin{align}
    \frac{1}{N}\mathrm{Tr}(B) &= 1 \,,\\
    \frac{1}{N}\mathrm{Tr}(B B^T\!) &= \beta \,,\\
    \frac{1}{N}\mathrm{Tr}(B B B^T\!) &= \beta^2 \,,\\
    \frac{1}{N}\mathrm{Tr}(B B^T\! B B^T\!) &= \gamma = \beta^4  \,,
\end{align}
with $B = (I - J)^{-1}$ and $\beta =  \frac{1}{1 - g^2}$. The matrix $J$ is a Gaussian random matrix whose entries are drawn independently from $\mathcal{N}(0, g^2 / N)$. We denote $W_0 = J$ in order to avoid the extra index. 

The traces generally stem from scalar products of the form $\mathbf{a}^T\! M \mathbf{a}$, where the entries of the random vector $\mathbf{a}$ are drawn from $\mathcal{N}(0, 1/N)$, and the matrix $M$ is independent of $\mathbf{a}$. In particular, any combinations of the matrices $B$ are independent of $\mathbf{a}$, since they only contain the random matrix $J$. Because of this independence, we have 
\begin{equation}
\label{eq:EaMa}
    \mathbb{E} \left[ \mathbf{a}^T\! M \mathbf{a}\right]
    = \sum_{i, j = 1}^{N} \mathbb{E} [ a_{i} M_{ij} a_{j}]
    = \sum_{i, j = 1}^{N} \underbrace{\mathbb{E} [ a_i a_j ] }_{ = \delta_{ij} / N}
    \mathbb{E} [ M_{ij}]
    = 
    \mathbb{E} \left[ \frac{\mathrm{Tr} M}{N} \right]
    \,.
\end{equation}
Computing the traces above and showing the self-averaging quality of the terms is a matter of counting the number of contributing combinations of $J$ and $J^T$. Our results are based on expanding $B$ into a geometric series
\begin{equation}
    B = I + \sum_{K=1}^{\infty} J^K  \,.
\end{equation}

\subsection{$\mathrm{Tr}(B)$}
We start with the trace of $B$ alone:
\begin{equation}
    \mathbb{E} \left[ \frac{\mathrm{Tr} B}{N} \right]
    = 1 + \sum_{K=1}^{\infty} 
    \frac{1}{N} \sum_{i=1}^N
    \mathbb{E}\left[ (J^K)_{ii}\right]
    = 1 
    + \mathcal{O}\left(\frac{1}{N}\right)
    \,.
\end{equation}
We show why the sum vanishes with $N$. For $K=1$, the entries $J_{ii}$ have expectation 0. For $K=2$, the independence of elements of $J$ yields 
\begin{equation}
    \frac{1}{N} \sum_{i=1}^N
    \mathbb{E}\left[ (J^2)_{ii}\right]
    = \frac{1}{N} \sum_{i, j=1}^N
    \mathbb{E}\left[ J_{ij} J_{ji}\right]
    = 
    \frac{1}{N} \sum_{i \ne j}
    \underbrace{\mathbb{E}\left[ J_{ij}\right] 
    \mathbb{E}\left[ J_{ji}\right]
    }_{=0}
    + 
    \frac{1}{N} \sum_{i} \underbrace{\mathbb{E}\left[ J_{ii}^2\right]}_{=g^2/N}
    = 
    \mathcal{O}\left(\frac{1}{N}\right)
    \,.
\end{equation}
The second term vanishes because there are only $N$ terms, but the factor $1/N$ before the sum and the contribution $g^2/N$ together yield $1/N^2$. This observation can be generalized to higher $K$:
\begin{equation}
    \frac{1}{N} \sum_{i=1}^N
    \mathbb{E}\left[ (J^K)_{ii}\right]
    = \frac{1}{N} \sum_{i_1, i_2, \dots, i_K}
    \mathbb{E}\left[ J_{i_1 i_2} J_{i_2 i_3} \dots J_{i_K i_1}\right]
    =
    \frac{1}{N} \sum_{i} \underbrace{\mathbb{E}\left[ J_{ii}^{K/2}\right]}_{= \mathcal{O}\left(N^{K/2}\right)}
    =
    \mathcal{O}\left(\frac{1}{N^{K/2-1}}\right)
    \,.
\end{equation}
There are $K$ different indices. Because each index appears once as a first and once as a second index, the attempt to form pairs directly results in setting all indices equal. 

\subsection{$\mathrm{Tr}(B B^T)$}
\label{sec:bbT}
The situation changes when introducing $B^T$. We can write 
\begin{equation}
    B B^T = \sum_{K, L=0}^\infty J^K J^{T L} \,,
\end{equation}
where the transpose $T$ and power $L$ commute. We compute the trace again term by term, starting at $K = L =1$:
\begin{equation}
    \frac{1}{N} \sum_{i=1}^N
    \mathbb{E}\left[ (J J^T)_{ii}\right]
    = \frac{1}{N} \sum_{i, j} 
    \mathbb{E}\left[ J_{ij} J^T_{ji}\right]
    =
    \frac{1}{N} \sum_{i, j} \underbrace{\mathbb{E}\left[ J_{ij}^2\right]}_{=g^{2}/N}
    = g^2 
    \,.
\end{equation}
For general $K, L \ge 1$, we have
\begin{equation}
\begin{split}
    \frac{1}{N} \sum_{i=1}^N
    \mathbb{E}\left[ (J^K J^{TL})_{ii}\right]
    &= \frac{1}{N} 
    \sum_{i_1, \dots i_K} 
    \sum_{j_1, \dots j_L} 
    \mathbb{E}\left[ 
    J_{i_1 i_2} J_{i_2 i_3} \dots J_{i_K j_1}
    J^T_{j_1 j_2} J^T_{j_2 j_3} \dots J^T_{j_L i_1}
    \right]
    \\&= \frac{1}{N} 
    \sum_{i_1, \dots i_K} 
    \sum_{j_1, \dots j_L} 
    \mathbb{E}\left[ 
    J_{i_1 i_2} J_{i_2 i_3} \dots J_{i_K j_1}
    J_{j_2 j_1} J_{j_3 j_2} \dots J^T_{j_1 i_L}
    \right]
    \,.
\end{split}
\end{equation}
We need to form pairs of indices. To simplify the discussion, we write the sequence of index pairs alone: 
\begin{equation}
    \begin{bmatrix}
    i_1 \\ i_2
    \end{bmatrix}
    \begin{bmatrix}
    i_2 \\ i_3
    \end{bmatrix}
    \dots
    \begin{bmatrix}
    i_{K-1} \\ i_K
    \end{bmatrix}
    \begin{bmatrix}
    i_K \\ j_1
    \end{bmatrix}
    \begin{bmatrix}
    j_2 \\ j_1
    \end{bmatrix}
    \begin{bmatrix}
    j_3 \\ j_2
    \end{bmatrix}
    \dots
    \begin{bmatrix}
    i_1 \\ j_L
    \end{bmatrix} \,.
\end{equation}
There are $K+L$ indices, and we need to form $(K+L)/2$ distinct pairs of index pairs. Each index constraint reduces the entire term by a factor of $1/N$. Because of the additional factor $1/N$ in front of the sum, we can have only $(K+L)/2 - 1$ index constraints. The question becomes one of counting the number of possible combinations. 

The expression above indicates that the only relevant term needs to constrain $i_K = j_2$. Under this condition, we have
\begin{equation}
    \begin{bmatrix}
    i_1 \\ i_2
    \end{bmatrix}
    \begin{bmatrix}
    i_2 \\ i_3
    \end{bmatrix}
    \dots
    \begin{bmatrix}
    i_{K-1} \\ i_K
    \end{bmatrix}
    \begin{bmatrix}
    i_K \\ j_1
    \end{bmatrix}
    \begin{bmatrix}
    i_K \\ j_1
    \end{bmatrix}
    \begin{bmatrix}
    j_3 \\ i_K
    \end{bmatrix}
    \dots
    \begin{bmatrix}
    i_1 \\ j_L
    \end{bmatrix} \,.
\end{equation}
The two middle terms drop and the new middle pairs show the same configuration. One can proceed iteratively with this scheme until reaching the right or left end (depending on $min(K, L)$). In fact, if $L > K$, then 
\begin{equation}
\underbrace{
    \begin{bmatrix}
    i_1 \\ i_2
    \end{bmatrix}
    \begin{bmatrix}
    i_2 \\ i_3
    \end{bmatrix}
    \dots
    \begin{bmatrix}
    i_{K-1} \\ i_K
    \end{bmatrix}
    \begin{bmatrix}
    i_K \\ j_1
    \end{bmatrix}
    \begin{bmatrix}
    i_K \\ j_1
    \end{bmatrix}
    \begin{bmatrix}
    i_{K-1} \\ i_K
    \end{bmatrix}
    \dots
    \begin{bmatrix}
    i_1 \\ i_2
    \end{bmatrix} 
    }_{\text{paired with $K-1$ constraints}}
    \begin{bmatrix}
    j_{K+1} \\ i_1
    \end{bmatrix} 
    \begin{bmatrix}
    j_{K+2} \\ j_{K+1}
    \end{bmatrix} 
    \dots
    \begin{bmatrix}
    i_1 \\ j_L
    \end{bmatrix} \,.
\end{equation}
The non-paired terms need $L-K$ additional constraints, so that the entire term only gives a contribution of $\mathcal{O}(1/N^{(L-K)/2 - 1})$. This and a similar argument for $K > L$ shows that we need $K=L$. In that case, there are $K-1 = (K+L)/2 - 1$ constraints and the term contributes at order $\mathcal{O}(1)$. We summarize with 
\begin{equation}
    \mathbb{E} \left[ \frac{\mathrm{Tr} ( J^K J^{T L} )}{N} \right]
    = g^{2K} \delta_{KL}  
    + \mathcal{O}\left(\frac{1}{N}\right)
    \,.
\end{equation}
For the entire matrix $B B^T$, this leads to 
\begin{equation}
    \mathbb{E} \left[ \frac{\mathrm{Tr} (B B^T) }{N} \right]
    = 
    \sum_{K, L=1}^{\infty} 
    \mathbb{E} \left[ \frac{\mathrm{Tr} (J^K J^{T L}) }{N} \right]
    =  
    \sum_{K=1}^{\infty} g^{2K} 
    + \mathcal{O}\left(\frac{1}{N}\right)
    = 
    \frac{1}{1 - g^2}
    + \mathcal{O}\left(\frac{1}{N}\right)
    \,.
\end{equation}
Note that the correction terms remain finite under the infinite sums for $K$ and $L$ because they scale with $g^{K+L}$ and we chose $g < 1$.

\subsection{$\mathrm{Tr}(B B B^T)$}
For $\mathrm{Tr}(B B B^T)$, the arguments go in parallel to the previous discussion. Indeed, we have 
\begin{equation}
\begin{split}
    \mathbb{E} \left[ \frac{\mathrm{Tr} (B B B^T)}{N} \right]
    &= 
    \sum_{K, L, M=1}^{\infty} 
    \mathbb{E} \left[ \frac{\mathrm{Tr} (J^K J^L J^{T M}) }{N} \right]
    \\&= 
    \sum_{K, L, M=1}^{\infty} g^{2M} \delta_{K+L, M} 
    \\&= 
    \sum_{M=0}^\infty g^{2M}
    \underbrace{\sum_{K=0}^\infty \sum_{L=0}^\infty \delta_{K+L, M}}_{=\sum_{K=0}^M 1}
    \\&= 
    \sum_{M=0}^\infty g^{2M} (M+1)
    \\&=
    \frac{1}{(1 - g^2)^2}
    \,.
\end{split}
\end{equation}
plus an order $\mathcal{O}(1/N)$ correction. 

\subsection{$\mathrm{Tr}(B B^T\! B B^T)$}
For $\mathbb{E}[\mathrm{Tr}(B B^T B B^T) / N]$, we first compute trace of the components 
$J^i J^{Tj} J^k J^{Tl}$. Similar to the cases discussed before, we need to constrain indices to create equal index pairs. The index pairs before any constraints can be written as
\setcounter{MaxMatrixCols}{20}
\begin{equation}
\label{eq:indices}
    \begin{bmatrix}
    i_1 & i_2 & \dots & i_i & j_2 & j_3 & \dots & k_1 & k_1 & k_2 & \dots & k_k & l_2 & l_3 & \dots & i_1 \\
    i_2 & i_3 & \dots & j_1 & j_1 & j_2 & \dots & j_j & k_2 & k_3 & \dots & l_1 & l_1 & l_2 & \dots & l_l
    \end{bmatrix} \,.
\end{equation}
There are $n=i+j+k+l$ summation indices, and each pair contributes with a factor $g^2/N$. Together with the additional factor $1/N$, we can thus have at most $n/2 - 1$ constraints. Note that like before, the number of transposed matrices must equal that of the non-transposed, $i+k = j+l$, so that $n$ is even. A smaller number of constraints is not sufficient, so that the question becomes: How many different sets of $n/2 - 1$ constraints lead to $n/2$ pairs of index pairs? 

We start with $i=j=k=l=1$. The corresponding index pairs are
\begin{equation}
    \begin{bmatrix}
    i_1 & k_1 & k_1 & i_1 \\
    j_1 & j_1 & l_1 & l_1
    \end{bmatrix} \,.
\end{equation}
One can see that there are two possible combinations to create two pairs:
$i_1 = k_1$ and $j_1 = l_1$, which yield
\begin{equation}
    \begin{bmatrix}
    i_1 & i_1 & i_1 & i_1 \\
    j_1 & j_1 & l_1 & l_1
    \end{bmatrix} \,,
    \qquad 
    \begin{bmatrix}
    i_1 & k_1 & k_1 & i_1 \\
    j_1 & j_1 & j_1 & j_1
    \end{bmatrix} \,.
\end{equation}
Therefore, there are 2 combinations. An index-counting argument like before generalizes this result, showing that the number of combinations is equal to
\begin{equation}
    c_{ijkl} = 1 + \min(i, j, k, l)\,.
\end{equation} 
We prove this statement by induction: Let $i = \min(i, j, k, l)$ without loss of generality (since the trace is cyclic). We rewrite the index pairs \cref{eq:indices} and color cases were two upper or lower indices are equal without any constraints: 
\setcounter{MaxMatrixCols}{20}
\begin{equation}
\label{eq:indices_c}
    \begin{bmatrix}
    \color{red} i_1 & i_2 & \dots & i_i & j_2 & j_3 & \dots & \color{green} k_1 & \color{green} k_1 & k_2 & \dots & k_k & l_2 & l_3 & \dots & \color{red} i_1 \\
    i_2 & i_3 & \dots & \color{blue} j_1 & \color{blue} j_1 & j_2 & \dots & j_j & k_2 & k_3 & \dots & \color{yellow} l_1 & \color{yellow} l_1 & l_2 & \dots & l_l
    \end{bmatrix} \,.
\end{equation}
We next separate two cases: Case 1, $i_i = j_2$, and Case 2, $i_i \ne j_2$. In Case 1, the index pairs with the blue $j$s above become equal:
\begin{equation}
    \begin{bmatrix}
    \color{red} i_1 & i_2 & \dots & i_{i-1 } & j_2 & j_2 & j_3 & \dots & \color{green} k_1 & \color{green} k_1 & k_2 & \dots & k_k & l_2 & l_3 & \dots & \color{red} i_1 \\
    i_2 & i_3 & \dots & j_2 & \color{blue} j_1 & \color{blue} j_1 & j_2 & \dots & j_j & k_2 & k_3 & \dots & \color{yellow} l_1 & \color{yellow} l_1 & l_2 & \dots & l_l
    \end{bmatrix} \,.
\end{equation}
We can take these pairs out, and the remaining indices read
\begin{equation}
    \begin{bmatrix}
    \color{red} i_1 & i_2 & \dots & i_{i-1 } & j_3 & j_4 & \dots & \color{green} k_1 & \color{green} k_1 & k_2 & \dots & k_k & l_2 & l_3 & \dots & \color{red} i_1 \\
    i_2 & i_3 & \dots & \color{blue} j_2 & \color{blue} j_2 & j_3 & \dots & j_j & k_2 & k_3 & \dots & \color{yellow} l_1 & \color{yellow} l_1 & l_2 & \dots & l_l
    \end{bmatrix} \,,
\end{equation}
where we colored the $j_2$ blue again.
We now have $(i', j', k', l') = (i-1, j-1, k, l)$ indices, with $\min(i', j', k', l') = i-1$. According to our induction hypothesis, there are $c_{i'j'k'l'} = 1 + i-1 = i$ different sets of $n/2 - 2$ constraints. Adding the constraint of Case 1, $i_i = j_2$ yields the expected number of $n/2 - 1$ constraints.

It remains to show that Case 2 allows for exactly one set of $n/2 - 1$ constraints. Because $i_i \ne j_2$ in \cref{eq:indices_c}, we need to have a pair at the red $i_1$; otherwise, one needs $n/2$ constraints. The pair at $i_1$ requires $l_l = i_2$, and dropping the newly formed pair yields
\begin{equation}
    \begin{bmatrix}
    \color{red} i_2 & i_3 & \dots & i_i & j_2 & j_3 & \dots & \color{green} k_1 & \color{green} k_1 & k_2 & \dots & k_k & l_2 & l_3 & \dots & \color{red} i_2 \\
    i_2 & i_3 & \dots & \color{blue} j_1 & \color{blue} j_1 & j_2 & \dots & j_j & k_2 & k_3 & \dots & \color{yellow} l_1 & \color{yellow} l_1 & l_2 & \dots & l_{l-1}
    \end{bmatrix} \,.
\end{equation}
We follow the same argumentation, constraining $l_{l-1} = i_3,\, \dots,\, l_{2 + l - i} = i_i$. We arrive at
\begin{equation}
    \begin{bmatrix}
    \color{red} i_i & j_2 & j_3 & \dots & \color{green} k_1 & \color{green} k_1 & k_2 & \dots & k_k & l_2 & \dots & \color{red} i_i \\
    \color{blue} j_1 & \color{blue} j_1 & j_2 & \dots & j_j & k_2 & k_3 & \dots & \color{yellow} l_1 & \color{yellow} l_1 & \dots & l_{1 + l - i}
    \end{bmatrix} \,.
\end{equation}
Further setting $l_{1 + l-i} = j_1$ and dropping the induced pair leads to
\begin{equation}
\label{eq:idxs_il}
    \begin{bmatrix}
    j_2 & j_3 & \dots & \color{green} k_1 & \color{green} k_1 & k_2 & \dots & k_k & l_2 & \dots & j_1 \\
    j_1 & j_2 & \dots & j_j & k_2 & k_3 & \dots & \color{yellow} l_1 & \color{yellow} l_1 & \dots & 
    l_{l - i}
    \end{bmatrix} \,.
\end{equation}
This is equal to the case $J^{Tj} J^k J^{T(l-i)}$. By the cyclic nature of the trace, this is equal to the case $J^{T(j+l-i)} J^{k}$. As discussed above, \cref{sec:bbT}, only one set of $(j + l - i + k)/2 - 1$ constraints leads to a full separation into pairs. Note that if $i = l$, the last set of indices, \cref{eq:idxs_il}, looks slightly different, but yields the same result.

Counting the number of constraints in Case 2 yields 
$1 + i-2 + 1 + (j + l - i + k) / 2 - 1 = n/2 - 1$. Since there is no other combination for Case 2, the total number of constraint combinations is precisely $i + 1 = 1 + \min(i,j,k,l) = c_{ijkl}$.

We return to the trace, which contains the factors $g^2$: 
\begin{equation}
    \mathbb{E} \left[ \frac{\mathrm{Tr} (J^i J^{Tj} J^k J^{Tl})}{N} \right]
    = g^{2(i+k)} \, \delta_{i+k, j+l} \, c_{ijkl} \,.
\end{equation}
We now evaluate the sums over $i, j, k, l$, starting with fixed $i$: 
\begin{equation}
    \sum_{j, k, l = 1}^{\infty}
    \mathbb{E} \left[ \frac{\mathrm{Tr} (J^i J^{Tj} J^k J^{Tl})}{N} \right]
    = 
    \sum_{j, k, l = 1}^{\infty}
    g^{2(i+k)} \, \delta_{i+k, j+l} \, c_{ijkl} \,.
\end{equation}
We split the summation into different regimes:
\begin{equation}
\begin{split}
    \sum_{j, k, l = 1}^{\infty}
    g^{2(i+k)} \, \delta_{i+k, j+l} \, c_{ijkl} 
    &=
    \sum_{\substack{j, l \\ j+l \ge i}}
    \sum_{k=1}^{\infty}
    g^{2(i+k)} \, \delta_{k, j+l - i} \, c_{ijkl} 
    \\&=
    \sum_{\substack{j, l \\ j \ge i \\ l \ge i}}
    g^{2(j+l)} \, (i+1)
    +
    \sum_{\substack{j, l \\ j + l \ge i \\ \min(j, l) < i}}
    g^{2(j+l)} \,  c_{ij (j+l-i) l} 
    \\&= a + b + c + d
    \,,
\end{split}
\end{equation}
where we split the second summand of the second-last line into two parts. The parts are:
\begin{align}
    a 
    &=
    \sum_{\substack{j, l \\ j \ge i \\ l \ge i}}
    g^{2(j+l)} \, (i+1)
    = (i+1) \sum_{j=i}^{\infty} (i+1) \left(\sum_{j=i}^\infty g^{2j}\right)^2
    = \frac{(i+1)g^{4i}}{(1 - g^2)^2} \,,
    \\ 
    b &= 
    \sum_{\substack{j, l \\ j < i \\ l < i \\ j+l \ge i}}
    g^{2(j+l)} \,  (j+l-i+1)
    = 
    \frac{g^{2i}}{(1 - g^2)^3} 
    \left[
    i(1+g^{2i})(1 - g^2) - (1 - g^{2i}) (1+ g^2)
    \right]  \,,
    \\
    c &= 
    \sum_{\substack{j, l \\ j \ge i \\ l < i}}
    g^{2(j+l)} \,  (l+1)
    = 
    \sum_{j=i}^{\infty} \sum_{l=0}^{i-1} g^{2l} \, (l+1)
    = 
    \frac{g^{2i}}{(1 - g^2)^3} 
    \left[
    1 - g^{2i} - g^{2i} i (1 - g^2)
    \right]  \,,
    \\
    d &= 
    \sum_{\substack{j, l \\ j < i \\ l \ge i}}
    g^{2(j+l)} \,  (j+1)
    = c \,.
\end{align}
Joining all terms yields
\begin{equation}
    \sum_{j, k, l = 1}^{\infty}
    \mathbb{E} \left[ \frac{\mathrm{Tr} (J^i J^{Tj} J^k J^{Tl})}{N} \right]
    = \frac{(i+1) g^{2i}}{(1 - g^2)^2} \,.
\end{equation}
Finally, we sum over $i$:
\begin{equation}
    \mathbb{E} \left[ \frac{\mathrm{Tr} (B B^T\! B B^T)}{N} \right]
    = 
    \sum_{i, j, k, l = 1}^{\infty}
    \mathbb{E} \left[ \frac{\mathrm{Tr} (J^i J^{Tj} J^k J^{Tl})}{N} \right]
    = \sum_{i=0}^\infty \frac{(i+1) g^{2i}}{(1 - g^2)^2} 
    = \frac{1}{(1 - g^2)^4} \,.
\end{equation}

We return to the trace, which is therefore
\begin{equation}
    \mathbb{E} \left[ \frac{\mathrm{Tr} (J^i J^{Tj} J^k J^{Tl})}{N} \right]
    = g^{2(i+k)} \, \delta_{i+k, j+l} \, c_{ijkl} \,.
\end{equation}
We now evaluate the sums over $i, j, k, l$, starting with fixed $i$: 
\begin{equation}
    \sum_{j, k, l = 1}^{\infty}
    \mathbb{E} \left[ \frac{\mathrm{Tr} (J^i J^{Tj} J^k J^{Tl})}{N} \right]
    = 
    \sum_{j, k, l = 1}^{\infty}
    g^{2(i+k)} \, \delta_{i+k, j+l} \, c_{ijkl} \,.
\end{equation}
We split the summation into different regimes:
\begin{equation}
\begin{split}
    \sum_{j, k, l = 1}^{\infty}
    g^{2(i+k)} \, \delta_{i+k, j+l} \, c_{ijkl} 
    &=
    \sum_{\substack{j, l \\ j+l \ge i}}
    \sum_{k=1}^{\infty}
    g^{2(i+k)} \, \delta_{k, j+l - i} \, c_{ijkl} 
    \\&=
    \sum_{\substack{j, l \\ j \ge i \\ l \ge i}}
    g^{2(j+l)} \, (i+1)
    +
    \sum_{\substack{j, l \\ j + l \ge i \\ \min(j, l) < i}}
    g^{2(j+l)} \,  c_{ij (j+l-i) l} 
    \\&= a + b + c + d
    \,,
\end{split}
\end{equation}
where we split the second summand of the second-last line into two parts. The parts are:
\begin{align}
    a 
    &=
    \sum_{\substack{j, l \\ j \ge i \\ l \ge i}}
    g^{2(j+l)} \, (i+1)
    = (i+1) \sum_{j=i}^{\infty} (i+1) \left(\sum_{j=i}^\infty g^{2j}\right)^2
    = \frac{(i+1)g^{4i}}{(1 - g^2)^2} \,,
    \\ 
    b &= 
    \sum_{\substack{j, l \\ j < i \\ l < i \\ j+l \ge i}}
    g^{2(j+l)} \,  (j+l-i+1)
    = 
    \frac{g^{2i}}{(1 - g^2)^3} 
    \left[
    i(1+g^{2i})(1 - g^2) - (1 - g^{2i}) (1+ g^2)
    \right]  \,,
    \\
    c &= 
    \sum_{\substack{j, l \\ j \ge i \\ l < i}}
    g^{2(j+l)} \,  (l+1)
    = 
    \sum_{j=i}^{\infty} \sum_{l=0}^{i-1} g^{2l} \, (l+1)
    = 
    \frac{g^{2i}}{(1 - g^2)^3} 
    \left[
    1 - g^{2i} - g^{2i} i (1 - g^2)
    \right]  \,,
    \\
    d &= 
    \sum_{\substack{j, l \\ j < i \\ l \ge i}}
    g^{2(j+l)} \,  (j+1)
    = c \,.
\end{align}
Joining all terms yields
\begin{equation}
    \sum_{j, k, l = 1}^{\infty}
    \mathbb{E} \left[ \frac{\mathrm{Tr} (J^i J^{Tj} J^k J^{Tl})}{N} \right]
    = \frac{(i+1) g^{2i}}{(1 - g^2)^2} \,.
\end{equation}
Finally, we sum over $i$:
\begin{equation}
    \mathbb{E} \left[ \frac{\mathrm{Tr} (B B^T\! B B^T)}{N} \right]
    = 
    \sum_{i, j, k, l = 1}^{\infty}
    \mathbb{E} \left[ \frac{\mathrm{Tr} (J^i J^{Tj} J^k J^{Tl})}{N} \right]
    = \sum_{i=0}^\infty \frac{(i+1) g^{2i}}{(1 - g^2)^2} 
    = \frac{1}{(1 - g^2)^4} \,,
\end{equation}
which is the statement we wanted to prove.

\section{Details of sentiment analysis task}
For the sentiment analysis task in the results section, we trained a 2-layer LSTM model on the Standford Sentiment Treebank with binary labels (SST-2) \citeSM{socher2013recursive}. The dataset consists of sentences from movie reviews which are labeled positive or negative. Sentences have on average 20 words, and there are 6920 training and 872 validation examples. We tokenized the sentences with the scaCy tokenizer \citeSM{spacy}. We further used a pretrained word embedding (GloVe, \citeSM{glove}) with dimension $N_\mathrm{in} = 100$. The word embedding was kept fixed during training.

Each LSTM layer had $N = 256$ units. All weights and biases were initialized from the uniform distribution $\mathcal{U}(-a, a)$, where $a = \sqrt{1 / N}$, except for input weights of layer 1, where $a = \sqrt{1/ N_\mathrm{in}}$. During training, all weights and biases were updated with Adam on a binary cross entropy loss, as implemented in PyTorch \citeSM{pytorch}. We set the learning rate to $0.01 / N$, and all other parameters at their default values. We additionally applied dropout with probability 0.5 to all hidden states. We trained the model for 500 epochs, each epoch iterating over the entire data set with batches of 64 sentences. 

To evaluate the performance after truncation, we separated the weights into recurrent and input weights. Because the LSTM for the four different gates are concatenated, the input weights of layer 1 have shape $4N \times N_\mathrm{in}$, all other weights have shape $4N \times N$. We simultaneously truncated the recurrent weights of both layers and the input weights of layer 2, i.e., all blocks with shape $4N \times N$ This specific choice did not alter the qualitative result, namely that truncating the changes $\Delta W$ and $\Delta U$ at a given rank produces a much smaller decrease in performance than truncating the full weights $W = W_0 + \Delta W$ and $U = U_0 + \Delta U$. 

Note that we chose the learning rate to be sufficiently small so that learning dynamics were smooth. With higher learning rates and rugged loss curves, we observed that changes $\Delta W$ would replace the initial connectivity, and the effective rank was much higher. 
Further note that other hyperparameters, such as L2 regularization on the weights, may also change the picture. 

\begin{figure}
\centering
\includegraphics[width=0.7\textwidth]{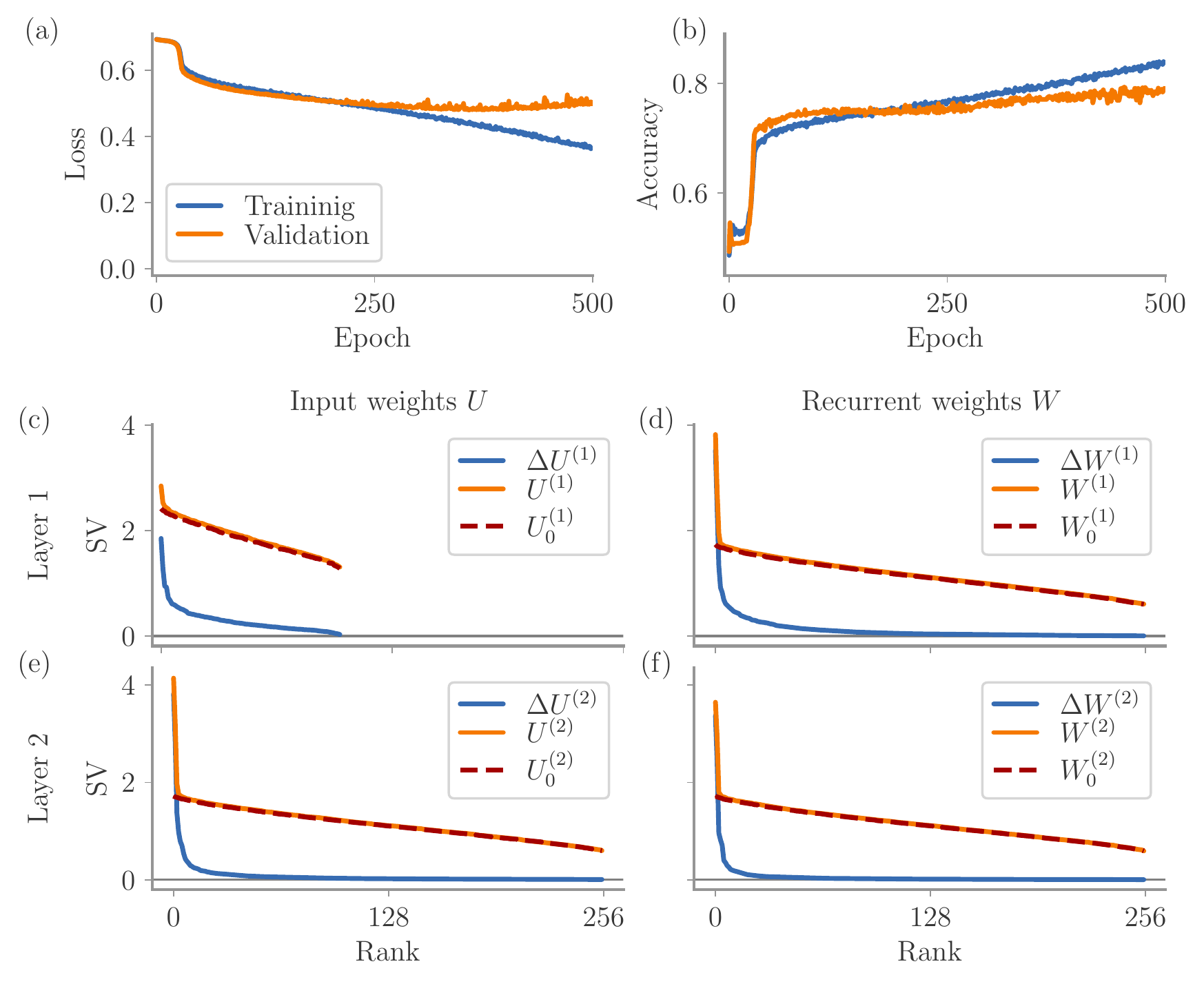}
\caption{
    Details for 
    2-layer LSTM model trained on a sentiment analysis task. 
    \textbf{(a, b)} Training and validation loss and accuracy over epochs. 
    \textbf{(c-f)} Singular values (SVs) of the input and recurrent weights in both layers. 
}
\label{fig:nlp_details}
\end{figure}

\bibliographystyleSM{plainnat}
\bibliographySM{neurips_2020_arxiv}

\end{document}